\documentclass[]{interact}
\usepackage{amsmath}
\usepackage{amssymb}
\usepackage{amsfonts}
\usepackage{mathtools}
\usepackage{bm}
\usepackage{color}
\usepackage{graphics}
\usepackage{graphicx}
\usepackage{epsfig}
\usepackage{epstopdf}
\usepackage[noadjust]{cite}
\usepackage{tikz}
\usepackage{pgfplots}
\usetikzlibrary{calc,positioning,shapes,shadows,arrows,fit}
\usepackage{fancyhdr}
\usepackage{fancyref}
\usepackage{hyperref}
\usepackage{float}
\usepackage[font = scriptsize]{caption}

\newcommand{\mat}[1]{\bm{\mathit{#1}}}
\newcommand{\vect}[1]{\bm{\mathbf{#1}}}

\newtheorem{theorem}{Theorem}
\newtheorem{lemma}{Lemma}

\theoremstyle{definition}
\newtheorem{definition}{Definition}

\newtheorem{assumption}{Assumption}
\theoremstyle{remark}
\newtheorem{remark}{Remark}

\newtheorem{property}{Property}
\theoremstyle{problem}
\newtheorem{problem}{Problem}

\begin{document}

\title{Robust Decentralized Navigation of Multi-Agent Systems with Collision Avoidance and Connectivity Maintenance Using Model Predictive Controllers}

\author{
\name{Alexandros Filotheou\textsuperscript{}, Alexandros Nikou\textsuperscript{}
and Dimos V. Dimarogonas\textsuperscript{}\thanks{CONTACT Alexandros Filotheou.
Email: alefil@kth.se, Alexandros Nikou. Email: anikou@kth.se and Dimos V.
Dimarogonas. Email: dimos@ee.kth.se}}
\affil{\textsuperscript{}The authors are with the KTH Center of Autonomous
Systems and ACCESS Linnaeus Center, School of Electrical Engineering and Computer Science, KTH Royal
Institute of Technology, SE-100 44, Stockholm, Sweden.}
}

\maketitle

\begin{abstract}
This paper addresses the problem of navigation control of a general class of
$2$nd order uncertain nonlinear multi-agent systems in a bounded workspace, which is a subset of
$\mathbb{R}^3$, with static obstacles. In particular, we propose a decentralized
control protocol such that each agent reaches a predefined position at the
workspace, while using local information based on a limited sensing radius.
The proposed scheme guarantees that the initially connected agents remain always
connected. In addition, by introducing certain distance constraints, we
guarantee inter-agent collision avoidance as well as collision avoidance with
the obstacles and the boundary of the workspace. The proposed controllers
employ a class of Decentralized Nonlinear Model Predictive Controllers
(DNMPC) under the presence of disturbances and uncertainties. Finally, simulation
results verify the validity of the proposed framework.
\end{abstract}


\section{Introduction}

During the last decades, decentralized control of multi-agent systems has
gained a significant amount of attention due to the great variety of its
applications, including  multi-robot systems, transportation, multi-point
surveillance and biological systems. The main focus of multi-agent systems is
the design of \emph{decentralized} control protocols in order to achieve global
tasks, such as \emph{consensus}
\cite{ren_beard_concensus, olfati_murray_concensus, jadbabaie_morse_coordination, tanner_flocking},
in which all the agents are required to converge to a specific point and
\emph{formation} \cite{egerstedt_formation, oh_park_ahn_2015, anderson_yu_fidan_hendrickx_2008, cao_morse_yu_anderson_dagsputa_2011, alex_cdc_2017_formation, alex_chris_ppc_formation_ifac, alex_automatica_formation},
in which all the agents aim to form a predefined geometrical shape. At the same
time, the agents might need to fulfill certain transient properties, such as
\emph{network connectivity}
\cite{magnus_2007_connectivity, zavlanos_2008_distributed, alex_cdc_2017_timed_abstractions} and/or
\emph{collision avoidance} \cite{dimos_2006_automatica_nf, alex_med_2017, ECC_2018_verginis}. In parallel,
another topic of research is \emph{multi-agent navigation} in both the
robotics and the control communities, due to the need for autonomous control
of multiple robotic agents in the same workspace. Important applications of
multi-agent navigation arise also in the fields of air-traffic management and in autonomous driving by guaranteeing
collision avoidance with other cars and obstacles. In this work, we study the
problem of multi-agent navigation with network connectivity and collision
avoidance properties.

The literature on approaching the problem of navigation of multi-agent systems
is rich. In \cite{dimos_2006_automatica_nf}, (\cite{makarem_decentralized_nf}),
a decentralized control protocol of multiple non-point agents (point masses)
with collision avoidance guarantees is considered. The problem is approached
by designing navigation functions which have been initially introduced in
\cite{koditschek1990robot}. A decentralized potential field
approach for navigation of multiple unicycles (aerial vehicles) with collision
avoidance has been considered in
\cite{panagou_potential_fields_unicycle, baras_decentralized_control}; Robustness analysis and saturation in control inputs are not
addressed. In \cite{roozbehani2009hamilton}, the collision avoidance problem
for multiple agents in intersections has been studied. An optimal control
problem is solved, with only time and energy constraints. Authors in
\cite{loizou_2017_navigation_transformation} proposed decentralized
controllers for multi-agent navigation and collision avoidance with
arbitrarily shaped obstacles in $2$D environments. Furthermore, connectivity maintenance properties are not taken into
consideration in all the aforementioned work.

Other approaches in multi-agent navigation propose solutions to decentralized
optimization problems. In \cite{Dunbar2006549}, a decentralized receding
horizon protocol for formation control of linear multi-agent systems is proposed.
Authors in \cite{aguiar_multiple_uavs_linearized_mpc} considered the
path-following problems for multiple Unmanned Aerial Vehicles (UAVs)  in which a decentralized optimization method is proposed through linearization of the
dynamics of the UAVs. A DNMPC along with potential functions for collision
avoidance has been studied in \cite{distributed_mpc_sastry}. A feedback
linearization framework along with Model Predictive Controllers (MPC) for
multiple unicycles in leader-follower networks for ensuring collision avoidance
and formation is introduced in \cite{fukushima_2005_distributed_mpc_linearization}.
Authors in \cite{kevicky1, kevicky2, 4459797} propose a decentralized receding horizon approach for discrete time multi-agent cooperative control.
However, in the aforementioned works, plant-model mismatch or uncertainties and/or connectivity maintenance are not considered.
In \cite{1383977} (\cite{1429425}), a centralized (decentralized) linear MPC
formulation and integer programming is proposed for dealing with collision
avoidance of multiple UAVs.

The contribution of this paper is to provide \emph{decentralized} control protocols which guarantee that a team of rigid-bodies modeled by $2$nd order \emph{uncertain} Lagrangian dynamics satisfy: collision avoidance between agents; obstacle avoidance; connectivity preservation; singularity avoidance; that agents remain in the workspace; while the control inputs are saturated. This constitutes a general problem that arises in many multi-agent applications where the agents need to perform a collaborative task, stay close and connected to each other and navigate to desired goal points. To the best of the authors' knowledge, decentralized control protocols that guarantee \emph{all} the aforementioned control specifications for the dynamics in hand have not been proposed in the bibliography. In order to address the aforementioned problem, we propose a Decentralized Nonlinear Model Predictive Control (DNMPC) framework in which each agent solves its own optimal control problem, having availability of information on the current and estimated actions of all agents within its sensing range. The proposed control scheme, under relatively standard Nonlinear Model Predictive Control (NMPC) assumptions, guarantees that all the aforementioned control specifications are satisfied. A conference version of this paper can be found in \cite{ecc_2018}, in which a similar problem is investigated for nonlinear uncertain dynamics with additive disturbance in $\mathbb{R}^n$, without any rotation representations. However, due to space constraints, the proofs have been omitted in the conference paper.

The remainder of this paper is structured as follows: In Section \ref{sec:notation_preliminaries} the notation and preliminaries background are given. Section \ref{sec:problem_formulation} provides the system dynamics and the formal problem statement. Section \ref{sec:main_results} discusses the technical details of the solution and Section \ref{sec:simulation_results} is devoted to simulation examples. Finally, conclusions and future work are discussed in Section \ref{sec:conclusions}.

\section{Notation and Preliminaries}
\label{sec:notation_preliminaries}
The set of positive integers is denoted by $\mathbb{N}$. The real $n$-coordinate
space, $n\in\mathbb{N}$, is denoted by $\mathbb{R}^n$;
$\mathbb{R}^n_{\geq 0}$ and $\mathbb{R}^n_{> 0}$ are the sets of real
$n$-vectors with all elements nonnegative and positive, respectively. Given a
set $S$, we denote by $\lvert S\lvert$ its cardinality. The notation
$\|\vect{x}\|$ is used for the Euclidean norm of a vector
$\vect{x} \in \mathbb{R}^n$ and
$\|\mat{A}\| = \max \{ \|\mat{A} \vect{x} \|: \|x\| = 1 \}$ for the induced norm
of a matrix $\mat{A} \in \mathbb{R}^{m \times n}$. Given a real symmetric matrix $\mat{A}$, $\lambda_{\text{min}}(\mat{A})$
and $\lambda_{\text{max}}(\mat{A})$ denote the minimum and the maximum absolute
value of eigenvalues of $\mat{A}$, respectively.
Its minimum and maximum singular values are denoted by
$\sigma_{\text{min}}(\mat{A})$ and $\sigma_{\text{max}}(\mat{A})$ respectively;
$\mat{I}_n \in \mathbb{R}^{n \times n}$ and
$\mat{0}_{m \times n} \in \mathbb{R}^{m \times n}$ are the unit matrix and the
$m \times n$ matrix with all entries zeros,
respectively. The set-valued function
$\mathcal{B}:\mathbb{R}^3\times\mathbb{R}_{> 0} \rightrightarrows \mathbb{R}^3$,
given by $\mathcal{B}(\vect{c}, r) = \{\vect{x} \in \mathbb{R}^3: \|\vect{x}-\vect{c}\| \leq r\},$
represents the $3$D sphere with center $\vect{c} \in \mathbb{R}^{3}$ and radius
$r \in \mathbb{R}_{> 0}$. Furthermore, we denote by $\phi$, $\theta$ and $\psi$
the Euler angles of a frame $\{\mathcal{F}\}$ with respect to an inertial frame
$\{\mathcal{F}_o\}$. We also use the notation $\mathcal{M} = \mathbb{R}^3\times \mathcal{T}^3$ where:
$\mathcal{T} = (-\pi, \pi) \times \left(-\frac{\pi}{2}, \frac{\pi}{2}\right) \times (\pi, \pi)$. For the definitions of Class $\mathcal{K}$, Class $\mathcal{KL}$ functions, Input-to-State Stability (ISS Stability), ISS Lyapunov Function and positively invariant sets, which will be used thereafter in this manuscript, we refer the reader to \cite{khalil_nonlinear_systems, marquez2003nonlinear, Sontag1995}.

\begin{definition} (\emph{Minkowski Addition}) \label{def:minkwoski}
	Given the sets $\mathcal{S}_1$, $\mathcal{S}_2 \subseteq \mathbb{R}^n$, their
	\emph{Minkowski addition} is defined by:
	$\mathcal{S}_1 \oplus \mathcal{S}_2 = \{s_1 + s_2 \ \in \mathbb{R}^n : s_1 \in \mathcal{S}_1, s_2 \in \mathcal{S}_2\}.$
\end{definition}
\begin{definition} (\emph{Pontryagin Difference}) \label{def:p_difference}
	Given the sets $\mathcal{S}_1$, $\mathcal{S}_2 \subseteq \mathbb{R}^n$, their \emph{Pontryagin difference} is defined by: $\mathcal{S}_1 \ominus \mathcal{S}_2 = \{s_1 \in \mathbb{R}^n: s_1+s_2 \in \mathcal{S}_1, \forall \ s_2 \in \mathcal{S}_2\}$.
\end{definition}

\begin{property} \label{property:set_property}
	Let the sets $\mathcal{S}_1$, $\mathcal{S}_2$, $\mathcal{S}_3 \subseteq \mathbb{R}^n$. Then, it holds that: $(\mathcal{S}_1 \ominus \mathcal{S}_2) \oplus (\mathcal{S}_2 \ominus \mathcal{S}_3) = (\mathcal{S}_1 \oplus \mathcal{S}_2) \ominus (\mathcal{S}_3 \oplus \mathcal{S}_3)$.
\end{property}
\begin{proof}
	The proof can be found in Appendix \ref{app:proof_set_property}.
\end{proof}

\section{Problem Formulation}
  \label{sec:problem_formulation}
  
\subsection{System Model}

Consider a set $\mathcal{V}$ of $N$ rigid bodies,
$\mathcal{V} = \{ 1,2, \ldots, N\}$, $N \geq 2$, operating in a workspace
$W\subseteq \mathbb{R}^3$. A coordinate frame $\{\mathcal{F}_i\}, i\in\mathcal{V}$
is attached to the center of mass of each body. The workspace is assumed to be
modeled as a bounded sphere
$\mathcal{B}\left(\vect{p}_{\scriptscriptstyle W},r_{\scriptscriptstyle W}\right)$
expressed in an inertial frame $\{\mathcal{F}_o\}$. We consider that over time $t$ each agent $i \in \mathcal{V}$ occupies the
space of a sphere $\mathcal{B}\left(\vect{p}_i(t), r_i\right)$, where
$\vect{p}_i : \mathbb{R}_{\geq 0} \to \mathbb{R}^3$
is the position of the agent's center of mass, and $r_i < r_W$ is the radius of the
agent's rigid body. We denote by $\vect{q}_i(t):\mathbb{R}_{\geq 0} \to \mathcal{T}^3$,
the Euler angles representing the agents' orientation with respect to the
inertial frame $\{\mathcal{F}_o\}$, with $\vect{q}_i \triangleq [\phi_i,\theta_i,\psi_i]^{\top}$.
By defining: $\vect{x}_i (t) \triangleq [\vect{p}_i(t)^{\top}$,
$\vect{q}_i(t)^{\top}]^{\top}$, $\vect{x}_i(t):\mathbb{R}_{\geq 0} \to \mathcal{M}$,
$\vect{v}_i(t) \triangleq [\dot{\vect{p}}_i(t)^{\top}$,
$\vect{\omega}_i(t)^{\top}]^{\top}$,
$\vect{v}_i(t) : \mathbb{R}_{\geq 0} \to \mathbb{R}^6$, we model the motion of
agent $i$ under continuous second order \emph{Lagrangian dynamics} as:
\begin{subequations}
	\begin{align}
	\dot{\vect{x}}_i(t) &= \mat{J}(\vect{q}_i) \vect{v}_i(t), \label{eq:system_1} \\
	\dot{\vect{v}}_i(t) &= \mat{M}^{-1}_i(\vect{x}_i) \left[-\mat{C}_i(\vect{x}_i,\dot{\vect{x}}_i) \vect{v}_i(t)-\vect{g}_i(\vect{x}_i) + \vect{u}_i(t) \right]+\widetilde{\vect{w}}_i(\vect{x}_i, \vect{v}_i, t), \label{eq:system_2}
	\end{align}
	\label{eq:system}
\end{subequations}
where $\mat{J}:\mathcal{T}^3 \to \mathbb{R}^{6\times6}$ is
a Jacobian matrix that maps the Euler angle rates to $\vect{v}_i$: $\mat{J}(\vect{q}_i) =
\begin{bmatrix}
\mat{I}_3 & \mat{0}_{3 \times 3} \\
\mat{0}_{3 \times 3} & \mat{J}_{ q}(\vect{q}_i) \\
\end{bmatrix}$,
$\mat{J}_q(\vect{q}_i) =
\begin{bmatrix}
1 & \sin\phi_i \tan\theta_i & \cos\phi_i \tan\theta_i \\
0 & \cos\phi_i & -\sin\phi_i \\
0 & \displaystyle \frac{\sin\phi_i}{\cos\theta_i} & \displaystyle \frac{\cos\phi_i}{\cos\theta_i}
\end{bmatrix}$. Moreover, $\mat{M}_i:\mathcal{M} \to \mathbb{R}^{6\times6}$ is the positive
definite \textit{inertia matrix},
$\mat{C}_i:\mathcal{M}\times\mathbb{R}^6 \to \mathbb{R}^{6\times6}$ is the
\textit{Coriolis matrix} and $\vect{g}_i:\mathcal{M} \to \mathbb{R}^6$ is the
\textit{gravity vector}. The continuous function
$\widetilde{\vect{w}}_i: \mathcal{M} \times \mathbb{R}^6 \times \mathbb{R}_{\ge 0} \to \mathbb{R}^{6}$
is a term representing \emph{disturbances} and \emph{modeling uncertainties}.
Finally, $\vect{u}_i:\mathbb{R}_{\ge 0}\to\mathbb{R}^6$ is the control
input vector representing the $6$D generalized \textit{actuation force} acting
on the agent. The aforementioned vectors as well as their derivatives are derived
with respect to the inertial frame $\mathcal{F}_o$. The matrix $\mat{J}(\vect{q}_i)$
is singular when $\cos\theta_i = 0$ $\Leftrightarrow$ $\theta_i$ $=$ $\pm \frac{\pi}{2}$.
However, the proposed controller guarantees that $\mat{J}(\vect{q}_i)$ is
well-defined for every $i \in \mathcal{V}$.

Let us define the vector
$\vect{z}_i(t) = \left[ \vect{x}_i(t)^\top, \vect{v}_i(t)^\top \right]^\top: \mathbb{R}_{\ge 0} \to \mathcal{M} \times \mathbb{R}^6$,
$i \in \mathcal{V}$. Then, by defining the vector
$\dot{\vect{z}}_i: \mathbb{R}_{\ge 0} \to \mathbb{R}^{12}$, the dynamics
\eqref{eq:system_1}, \eqref{eq:system_2} can be written as:
\begin{equation} \label{eq:system_perturbed}
\dot{\vect{z}}_i(t) = f_i(\vect{z}_i(t), \vect{u}_i(t)) + \vect{w}_i(\vect{z}_i(t), t),
\end{equation}
where $\vect{w} = \left[\mat{0}_{1 \times 6}, \widetilde{w}_i^\top \right]^\top$
and the functions $f_i: \mathcal{M} \times \mathbb{R}^6 \times \mathbb{R}^6 \to \mathbb{R}^{12}$, $i \in \mathcal{V}$ are given by: $f_i(\vect{z}_i(t), \vect{u}_i(t)) \ \triangleq \
\begin{bmatrix}
\mat{J} \vect{v}_i(t) \\
-\mat{M}_i^{-1} \left[ \mat{C}_i \vect{v}_i(t) + \vect{g}_i - \vect{u}_i(t) \right]
\end{bmatrix}$. It is assumed that there exist finite constants $\bar{w}_i$, $\bar{u}_i \in \mathbb{R}_{> 0}$, $i \in \mathcal{V}$ such that:
\begin{align}
\mathcal{W}_i = \{ \vect{w}_i \in \mathbb{R}^{12} : \|\vect{w}_i\| \leq \overline{w}_i\}, \mathcal{U}_i = \{ \vect{u}_i \in \mathbb{R}^6 : \|\vect{u}_i\| \leq \overline{u}_i\}, \label{eq:mathcal_WU}
\end{align}
i.e., the disturbances $w_i$ as well as the control inputs $u_i$ are upper
bounded by the terms $\overline{w}_i$, $\overline{u}_i$, respectively.

\begin{assumption}
	\label{ass:g_i_g_R_Lipschitz}
	The nonlinear functions $f_i$ are \emph{locally Lipschitz continuous} in $\mathcal{M} \times \mathbb{R}^6 \times \mathcal{U}_i$ with Lipschitz constants $L_{f_i}$. Thus, it holds that:
	\begin{align} \label{eq:lipschitz_f_i}
	\| f_i(\vect{z}, \vect{u}) - f_i(\vect{z}', \vect{u}) \| \le L_{f_i} \| \vect{z} - \vect{z}' \|, \forall \vect{z}, \vect{z}' \in \mathcal{M} \times \mathbb{R}^6, \vect{u} \in \mathcal{U}_i.
	\end{align}
\end{assumption}

We consider that in the given workspace there exists a set of
$L \in \mathbb{N}$ \emph{static obstacles}, with $\mathcal{L} = \{1, 2, \dots, L\}$,
also modeled by the spheres
$\mathcal{B}\left(\vect{p}_{\scriptscriptstyle O_\ell}, r_{\scriptscriptstyle O_\ell}\right)$,
with centers at positions $\vect{p}_{\scriptscriptstyle O_\ell} \in \mathbb{R}^3$
and radii $r_{\scriptscriptstyle O_\ell}\in \mathbb{R}_{> 0}$,
where $\ell \in \mathcal{L}$. Their position and size in the 3D space is assumed to
	be a priori unknown to each agent. In order for agents to be able
	to detect the obstacles during their navigation, we assume that each agent
	$i \in \mathcal{V}$ has a limited spatial obstacle-detection range $b_i$ such
	that $b_i > r_i$. Thus, each agent senses points which reside on the surface of the
	obstacles and which are within a radius $b_i$ of its position. Given these
	points, each agent reconstructs the sphere that corresponds to the obstacle and
	extracts its position and radius in $3$D space.

\begin{assumption} (\emph{Measurements Assumption})
	\label{ass:measurements_access} Agent $i \in\mathcal{V}$ has: $1)$ access
	to measurements $\vect{p}_i, \vect{q}_i, \dot{\vect{p}}_i, \vect{\omega}_i,$
	that is, vectors $\vect{x}_i, \vect{v}_i$ pertaining to itself;
	$2)$ A limited sensing range $d_i$ such that:
	$\displaystyle d_i > \max_{i,j \in \mathcal{V}, i \neq j}\{r_i + r_j\}.$
\end{assumption}

\begin{figure}[t!]
	\centering
	\begin{tikzpicture}[scale = 0.4]
	\draw [color=black,thick,->,>=stealth'](-9, -5) to (-7, -5);
	\draw [color=black,thick,->,>=stealth'](-9, -5) to (-9, -3);
	\draw [color=black,thick,->,>=stealth'](-9, -5) to (-10, -6.5);
	\node at (-9.8, -5.0) {$\{\mathcal{F}_o\}$};
	
	\draw [color = blue, fill = blue!20] (-4.5,0) circle (2.5cm);
	\draw [green,thick,dashed] (-4.5,0) circle (5.0cm);
	\draw [blue,thick,dashed] (-4.5,0) circle (8.0cm);
	\draw [color=blue,thick,dashed,->,>=stealth'](-4.5, 0.0) to (2.0532, 4.5886);
	\draw [color=black,thick,->,>=stealth'](-9, -5) to (-4.5, -0.1);
	\node at (-7.80, -2.6) {$\vect{p}_i(\tau)$};
	\draw [color=green,thick,dashed,->,>=stealth'](-4.5, 0.0) to (-8.93, 2.43);
	\node at (-7.3, 2.15) {$d_i$};
	\node at (-1.8378,2.4641) {$b_i$};
	\draw [color=black,thick,dashed,->,>=stealth'](-4.5, 0.0) to (-2.0, 0.0);
	\node at (-3.3, 0.3) {$r_i$};
	\node at (-5.5, 0.0) {$\{\mathcal{F}_i\}$};
	\node at (-4.5, 0.0) {$\bullet$};
	\node at (-4.8, 3.0) {$\text{Agent} \ i$};
	
	\draw [color = red, fill = red!20] (8.2, 0) circle (1.5cm);
	\draw[orange,thick,dashed,] (8.2, 0) circle (4.1cm);
	\draw [color=black,thick,->,>=stealth'](-9, -5) to (8.2, -0.1);
	\node at (-3.0, -4.0) {$\vect{p}_j(\tau)$};
	\draw [color=orange,thick,dashed,->,>=stealth'](8.2, 0.0) to (8.2, -4.0);
	\node at (8.8, -2.7) {$d_j$};
	\draw [color=black,thick,dashed,->,>=stealth'](8.2, 0.0) to (4.1, 0.0);
	\node at (4.6, 0.4) {$r_j$};
	\node at (8.2, 0.0) {$\bullet$};
	\node at (8.0, 2.1) {$\text{Agent} \ j$};
	\node at (9.5, 0.0) {$\{\mathcal{F}_j\}$};
	
	\draw [color = black, fill = black!20] (-1, -8) circle (1.2cm);
	\draw [color=black,thick,->,>=stealth'](-9, -5) to (-1.1, -7.98);
	\draw [color=black,thick,dashed,->,>=stealth'](-1, -8) to (-1, -6.8);
	\node at (-1, -8) {$\bullet$};
	\node at (-5.0, -6.0) {$\vect{p}_{\scriptscriptstyle O_\ell}$};
	\node at (2.3, -8.0) {$\text{Obstacle}$};
	\node at (-0.40, -7.5) {$r_{\scriptscriptstyle O_\ell}$};
	\draw  [color=blue, very thick](-1,-8) ++(65:1.2) arc (65:160:1.2);
	\end{tikzpicture}
	\caption{Illustration of two agents $i, j \in \mathcal{V}$ and a static
		obstacle $\ell \in \mathcal{L}$ in the workspace at a time instant $\tau$;
		$\{\mathcal{O}\}$ is the inertial frame, $\{\mathcal{F}_i\}$,
		$\{\mathcal{F}_j\}$ are the frames attached to the agents' center of mass,
		$\vect{p}_i, \vect{p}_j, \vect{p}_{\ell} \in \mathbb{R}^3$ are the
		positions of the centers of mass of agents $i,j$ and obstacle $\ell$
		respectively, expressed in frame $\{\mathcal{F}_o\}$; $r_i, r_j, r_{\ell}$
		are the radii of the agents $i,j$ and the obstacle $\ell$ respectively;
		$d_i, d_j$ with $d_i > d_j$ are the agents' sensing ranges; $b_i$ is
			the spatial obstacle-detection range of agent $i$.}
	\label{fig:two_agents_one_obstacle}
\end{figure}
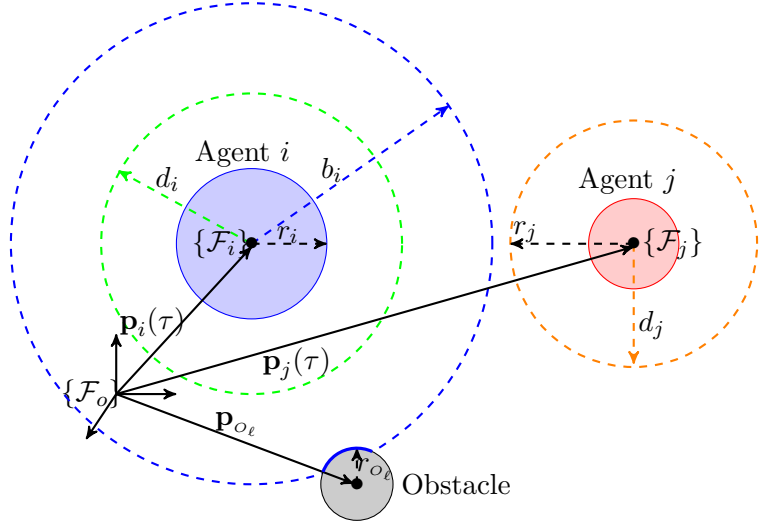

The latter implies that each agent has sufficiently large sensing radius so as to measure the agent with the biggest volume, due to the fact that the agents' radii are not the same. The consequence of points 1 and 2 of Assumption \ref{ass:measurements_access} is that by defining the set of agents $j$ that are
within the sensing range of agent $i$ at time $t$ as:
$\mathcal{R}_i(t) \triangleq \{j\in\mathcal{V} \backslash \{i\} : \| \vect{p}_i(t) - \vect{p}_j(t) \| < d_i\},$
agent $i$ knows all signals $\vect{p}_{j}(t)$, $\vect{q}_{j}(t)$,
$\dot{\vect{p}}_j(t)$, $\vect{\omega}_j(t)$, $\forall j\in \mathcal{R}_i(t)$,
$t\in\mathbb{R}_{\geq 0}$, of all agents $j \in \mathcal{R}_i(t)$ by virtue of
being able to calculate them using knowledge of its own $\vect{p}_i(t)$,
$\vect{q}_i(t)$, $\dot{\vect{p}}_i(t)$, $\vect{\omega}_i(t)$. The geometry of
two agents $i$ and $j$ as well as an obstacle $\ell$ in the workspace $W$ is
depicted in Figure \ref{fig:two_agents_one_obstacle}.

\begin{definition} (\textit{Collision/Singularity-free Configuration}) \label{definition:collision_free_conf}
	The multi-agent system is in a collision/singularity-free configuration at a time instant $\tau \in \mathbb{R}_{\ge 0}$ if all the following hold: $1)$ For every $i$, $j \in \mathcal{V}$, $i \neq j$ it holds that: $\| \vect{p}_i(\tau) - \vect{p}_j(\tau) \| > r_{i} + r_{j}$; $2)$ For every $i \in \mathcal{V}$ and for every $\ell \in \mathcal{L}$ it holds that: $\| \vect{p}_i(\tau) - \vect{p}_{\scriptscriptstyle O_\ell} \| > r_{i} + r_{\scriptscriptstyle O_\ell}$; $3)$ For every $i \in \mathcal{V}$ it holds that: $\|\vect{p}_i(\tau)-\vect{p}_{\scriptscriptstyle W}\| < r_{\scriptscriptstyle W} - r_i$; $4)$ For every $i \in \mathcal{V}$ it holds that: $-\frac{\pi}{2} < \theta_i(\tau) < \frac{\pi}{2}$.
\end{definition}

\begin{definition} (\textit{Neighboring set})
	Define the \emph{neighboring set} of agent $i \in \mathcal{V}$ as: $\mathcal{N}_i = \{j \in \mathcal{V} \backslash \{i\} : j \in \mathcal{R}_i(0) \}$. We will refer to agents $j \in \mathcal{N}_i$ as the \textit{neighbors} of agent $i \in \mathcal{V}$.
\end{definition}

The set $\mathcal{N}_i$ is composed of indices of agents
$j \in \mathcal{V}$ which are within the sensing range of agent $i$ at time
$t=0$. Agents $j \in \mathcal{N}_i$ are agents which agent $i$ is instructed
to keep within its sensing range at all times $t \in \mathbb{R}_{>0}$, and therefore
maintain connectivity with. While the sets $\mathcal{N}_i$ are introduced
for connectivity maintenance specifications and they are fixed, the sets $\mathcal{R}_i(t)$ are used to ensure collision avoidance, and, in general,
their composition evolves and varies through time.

\begin{assumption} (\emph{Initial Conditions Assumption}) \label{ass:initial_conditions}
	For sake of cooperation needs, we assume that $\mathcal{N}_i \neq \emptyset$, $\forall i \in \mathcal{V}$ i.e., all agents
	have at least one neighbor. We also assume that at time $t=0$ it holds that
	$\vect{v}_i(0) = \mat{0}_{6 \times 1}$ and the multi-agent system is in a
	\textit{collision/singularity-free configuration}, as per Definition
	\ref{definition:collision_free_conf}.
\end{assumption}

\subsection{Objectives}

Given the aforementioned modeling of the system, the objective of this paper is the \textit{stabilization of the agents} $i \in \mathcal{V}$ starting
from a collision/singularity-free configuration as given in Definition \ref{definition:collision_free_conf} to a desired feasible configuration
$\vect{x}_{i, \text{des}} = [\vect{p}_{i, \text{des}}^{\top},\vect{q}_{i, \text{des}}^{\top}]^{\top} \in \mathcal{M}$, while maintaining connectivity between neighboring agents, and avoiding collisions between agents, obstacles, and the workspace boundary.

\begin{definition} (\textit{Desired Feasible Configuration}) \label{definition:feasible_steady_state_conf}
	The desired configuration $\vect{x}_{i, \text{des}} = [\vect{p}_{i, \text{des}}^{\top},\vect{q}_{i, \text{des}}^{\top}]^{\top} \in \mathcal{M}$ of agents $i \in \mathcal{V}, j \in \mathcal{N}_i$ is \textit{feasible} if the following hold: $1)$ It is a collision/singularity-free configuration according to Definition \ref{definition:collision_free_conf}; $2)$ It does not result in a violation of the connectivity maintenance between neighboring agents, i.e., $\|\vect{p}_{i, \text{des}} - \vect{p}_{j, \text{des}}\| < d_i$, $\forall i \in \mathcal{V}, j \in \mathcal{N}_i$.
\end{definition}

\begin{definition} (\emph{Feasible Initial Conditions}) \label{def:set_feasible_initial_conditions}
	Let $\vect{x}_{i, \text{des}} = \left[ \vect{p}_{i, \text{des}}^\top, \vect{q}_{i, \text{des}}^\top \right]^{\top} \in \mathcal{M}$, $i \in \mathcal{V}$ be a desired feasible configuration as defined in Definition \ref{definition:feasible_steady_state_conf}. Then, the set of all initial conditions $\vect{x}_i(0)$, $\vect{v}_i(0)$ according to Assumption \ref{ass:initial_conditions}, for which there exist time constants $\overline{t}_i \in \mathbb{R}_{> 0} \cup \{\infty\}$ and control inputs $\vect{u}^\star_i \in \mathcal{U}_i$, $i \in \mathcal{V}$, which define a solution $\vect{x}_i^\star(t)$, $t \in [0, \overline{t}_i]$ of the system of differential equations \eqref{eq:system_1}-\eqref{eq:system_2}, under the presence of disturbance $w_i \in \mathcal{W}_i$, such that: $1)$ $\vect{x}_i^\star(\overline{t}_i) = \vect{x}_{i, \text{des}}$, $2)$ $\| \vect{p}_i^\star(t) - \vect{p}_j^\star(t) \| > r_{i} + r_{j}$ for every $t \in [0, \overline{t}_i]$, $i, j \in \mathcal{V}$, $i \neq j$, $3)$ $\|\vect{p}^\star_i(t) - \vect{p}_{\scriptscriptstyle O_\ell}\| > r_{i} + r_{\scriptscriptstyle O_\ell}$ for every $t \in [0, \overline{t}_i]$, $i \in \mathcal{V}$, $\ell \in \mathcal{L}$, $4)$ $\|\vect{p}^\star_i(t)-\vect{p}_{\scriptscriptstyle W}\| < r_{\scriptscriptstyle W} - r_i$ for every $t \in [0, \overline{t}_i]$, $i \in \mathcal{V}$, $5)$ $\|\vect{p}^\star_i(t) - \vect{p}^\star_j(t)\| < d_i$ for every $t \in [0, \overline{t}_i]$, $i \in \mathcal{V}, j \in \mathcal{N}_i$, are called \emph{feasible initial conditions}.
\end{definition}
The feasible initial conditions are, essentially, all the initial conditions $\vect{x}_i(0)$, $\vect{v}_i(0)$, $i \in \mathcal{V}$ from which there exist controllers $u_i \in \mathcal{U}_i$ that can navigate the agents to the given desired states $\vect{x}_{i, \text{des}}$, under the presence of disturbances $w_i \in \mathcal{W}_i$, while i) the initial neighbors remain connected, ii) the agents do not collide with each other, iii) the agents stay in the workspace and iv) the agents do to collide with the obstacles of the environment. Initial conditions for which one or more agents can not be driven to the desired state $\vect{x}_{i, \text{des}}$ by a controller $u_i \in \mathcal{U}_i$, i.e., initial conditions that violate one or more of the conditions of Definition \ref{def:set_feasible_initial_conditions}, are considered \emph{infeasible initial conditions}. 
Motivated by this observation, the goal of this paper is to provide a systematic method of designing decentralized feedback controllers that navigate the agents to the desired states $\vect{x}_{i, \text{des}}$ from all \emph{feasible initial conditions}, as defined in Definition \ref{def:set_feasible_initial_conditions}.

\subsection{Problem Statement}
\label{sec:problem_statement}

Formally, the control problem, under the aforementioned constraints, is
formulated as follows:
\begin{problem} \label{problem}
	Consider $N$ agents governed by dynamics \eqref{eq:system_perturbed},
	modeled by the spheres $\mathcal{B}\left(\vect{p}_i, r_i\right)$,
	$i \in \mathcal{V}$, and operating in a spherical workspace $W$ which is
	modeled by the sphere
	$\mathcal{B}\left(\vect{p}_{\scriptscriptstyle W}, r_{\scriptscriptstyle W}\right)$. In the workspace there are $L$ spherical obstacles $\mathcal{B}\left(\vect{p}_{\scriptscriptstyle O_\ell}, r_{\scriptscriptstyle O_\ell}\right)$,
	$\ell \in \mathcal{L}$. The agents have communication capabilities according
	to Assumption \ref{ass:measurements_access}, under the initial conditions
	$\vect{x}_i(0)$, $\vect{v}_i(0)$ imposed by Assumption
	\ref{ass:initial_conditions} and they are affected by disturbances
	$w_i \in \mathcal{W}_i$. Then, given a desired feasible configuration
	$\vect{x}_{i, \text{des}}$ according to Definition
	\ref{definition:feasible_steady_state_conf}, for all feasible initial
	conditions, as defined in Definition \ref{def:set_feasible_initial_conditions}, the problem lies in designing \emph{decentralized feedback control} laws $\vect{u}_i \in \mathcal{U}_i$,
	such that for every $i \in \mathcal{V}$ and for all times
	$t \in \mathbb{R}_{\geq 0}$, all the following specifications are satisfied: $1)$ Position and orientation stabilization is achieved: $\lim_{t \to \infty} \|\vect{x}_i(t) - \vect{x}_{i, \text{des}} \| \to 0;$ $2)$ Inter-agent collision is avoided: $\|\vect{p}_i(t) - \vect{p}_j(t)\| > r_{i} + r_{j}, \forall j \in \mathcal{V}, j \neq i;$ $3)$ Connectivity between neighboring agents is preserved: $\|\vect{p}_i(t) - \vect{p}_j(t)\| < d_i, \forall j \in \mathcal{N}_i;$ $4)$ Agent-with-obstacle collision is avoided: $\|\vect{p}_i(t) - \vect{p}_{\scriptscriptstyle O_\ell}(t)\| > r_{i} + r_{\scriptscriptstyle O_\ell}, \forall \ell \in \mathcal{L};$ $5)$ Agent-with-workspace-boundary collision is avoided: $\|\vect{p}_i(t)-\vect{p}_{\scriptscriptstyle W}\| < r_{\scriptscriptstyle W} - r_i;$ $6)$ All matrices $\mat{J}(\vect{q}_i)$ are well defined: $- \frac{\pi}{2} < \theta_i(t) < \frac{\pi}{2};$
\end{problem}

\section{Proposed Solution}
  \label{sec:main_results}
In this section, a systematic solution to Problem \ref{problem} is introduced.
Our overall approach builds on designing a decentralized control law
$\vect{u}_i \in \mathcal{U}_i$, $i \in \mathcal{V}$ for each agent. In
particular, since we aim to minimize the norms
$\|\vect{x}_i(t) - \vect{x}_{i, \text{des}} \|$ as $t \to \infty$, subject to
the state constraints imposed by Problem \ref{problem}, it is reasonable to seek
a solution which is the outcome of an optimization problem. In Section
\ref{sec:error_dynamics} we derive the error dynamics and in Section
\ref{sec:control_design} we discuss the proposed control scheme as well as the
stability analysis.

\subsection{Error Dynamics} \label{sec:error_dynamics}

Let us define the stack vector of the desired states and velocities by:
$\vect{z}_{i, \text{des}} = \left[ \vect{x}_{i, \text{des}}^\top, \vect{v}^\top_{i, \text{des}} \right]^\top \in \mathcal{M} \times \mathbb{R}^6$.
The state $\vect{x}_{i, \text{des}} \in \mathcal{M}$ is the desired feasible
state that agent $i$ needs to reach, as is given in Problem \ref{problem}.
For the desired velocities $\vect{v}_{i, \text{des}} \in \mathbb{R}^6$ we can
set, without loss of generality, that $\vect{v}_{i, \text{des}} = \mat{0}_{6 \times 1}$,
i.e., the agents need to stop when they achieve the desired state. We define
the error vector $\vect{e}_i: \mathbb{R}_{\ge 0} \to \mathcal{M} \times \mathbb{R}^6$ by:
\begin{equation}\label{eq:error_def}
\vect{e}_i(t)
=
\begin{bmatrix}
\vect{x}_{i}(t) \\
\vect{v}_{i}(t) \\
\end{bmatrix}
-
\begin{bmatrix}
\vect{x}_{i, \text{des}} \\
\vect{v}_{i, \text{des}} \\
\end{bmatrix}
=
\vect{z}_i(t)-\vect{z}_{i, \text{des}}.
\end{equation}
If we provide a control scheme that guarantees that
$\displaystyle \lim_{t \to \infty} \|\vect{z}_i(t)-\vect{z}_{i, \text{des}}\| \to 0$
then it is also guaranteed that $\displaystyle \lim_{t \to \infty} \|\vect{x}_i(t)-\vect{x}_{i, \text{des}}\| \to 0$,
which is the first goal of Problem \ref{problem}. By defining the
vector $\dot{\vect{e}}_i: \mathbb{R}_{\ge 0} \to \mathbb{R}^{12}$, the
\emph{error dynamics} are given by:
\begin{equation} \label{eq:error_system_perturbed}
\dot{\vect{e}}_i(t) = h_i(\vect{e}_i(t), \vect{u}_i(t)),
\end{equation}
where the functions
$h_i: \mathcal{M} \times \mathbb{R}^6 \times \mathbb{R}^6 \to \mathbb{R}^{12}$, $g_i: \mathcal{M} \times \mathbb{R}^6 \times \mathbb{R}^6 \to \mathbb{R}^{12}$ are defined by:
\begin{subequations}
	\begin{align}
	h_i(\vect{e}_i(t), \vect{u}_i(t)) & \triangleq g_i(\vect{e}_i(t), \vect{u}_i(t)) + \vect{w}_i(\vect{e}_i(t)+\vect{z}_{i, \text{des}}, t), \label{eq:functions_h_i} \\
	g_i(\vect{e}_i(t), \vect{u}_i(t)) & \triangleq f_i(\vect{e}_i(t)+\vect{z}_{i, \text{des}}, \vect{u}_i(t)), \label{eq:functions_g_i}
	\end{align}
\end{subequations}
respectively, where $f_i$ is defined in \eqref{eq:system_perturbed}. We define
the set $\mathcal{Z}_{i} \subseteq \mathcal{M} \times \mathbb{R}^{6}$, $i \in \mathcal{V}$
as the set that captures all the \textit{state} constraints on the system
\eqref{eq:system}, posed by Problem \ref{problem}. Therefore $\mathcal{Z}_{i}$
is given by:
\begin{align*}
\mathcal{Z}_{i} \triangleq & \Big\{ \vect{z}_i(t) \in \mathcal{M} \times \mathbb{R}^6 : \
\|\vect{p}_i(t) - \vect{p}_j(t)\| \ge r_i+r_j+\varepsilon, \forall j \in \mathcal{R}_i(t), \notag \\ 
&\|\vect{p}_i(t) - \vect{p}_j(t)\| \le d_i - \varepsilon, \forall j \in \mathcal{N}_i, \|\vect{p}_i(t) - \vect{p}_{\scriptscriptstyle O_\ell}\| \ge r_{i} + r_{\scriptscriptstyle O_\ell} + \varepsilon, \forall \ell \in \mathcal{L}, \\
& \|\vect{p}_i(t)-\vect{p}_{\scriptscriptstyle W}\| \le r_{\scriptscriptstyle W} - r_i - \varepsilon, - \frac{\pi}{2} + \varepsilon \le \theta_i(t) \le  \frac{\pi}{2} - \varepsilon \Big\}, i \in \mathcal{V},
\end{align*}
where $\varepsilon \in \mathbb{R}_{> 0}$ is an arbitrary small constant. In
order to translate the constraints that are dictated for the state $z_i$ into
constraints regarding the error state $e_i$ of \eqref{eq:error_def}, we define
the set $\mathcal{E}_{i} = \left\{\vect{e}_i \in \mathcal{M} \times \mathbb{R}^6 : \vect{e}_i \in \mathcal{Z}_{i} \oplus (-\vect{z}_{i, \text{des}}) \right\}$, $\forall i \in \mathcal{V}$. Then, the following rudimentary equivalence holds for all
$i \in \mathcal{V}$: $\vect{z}_i \in \mathcal{Z}_i \Leftrightarrow \vect{e}_i \in \mathcal{E}_i$.

\begin{property} \label{property 1}
	The nonlinear functions $g_i$, $i \in \mathcal{V}$ as defined
	in \eqref{eq:functions_g_i}, are \emph{locally Lipschitz continuous} in $\mathcal{E}_i \times \mathcal{U}_i$, with
	Lipschitz constants $L_{g_i} = L_{f_i}$, where $L_{f_i}$ as in
	\eqref{eq:lipschitz_f_i}. Thus,
	\begin{equation} \label{eq:lipsitch_g_i}
	\|g_i(\vect{e}, \vect{u})-g_i(\vect{e}', \vect{u})\| \le L_{g_i} \|\vect{e}-\vect{e}'\|, \forall \vect{e}, \vect{e}' \in \mathcal{E}_i, \vect{u} \in \mathcal{U}_i.
	\end{equation}
	\begin{proof}
		The proof can be found in Appendix \ref{app:proof_property_1}.
	\end{proof}
\end{property}

The goal is to solve Problem \ref{problem}, i.e, to design decentralized
control laws $\vect{u}_i \in \mathcal{U}_i$, $\forall i \in \mathcal{V}$ such
that the error signal $\vect{e}_i$, with dynamics as in \eqref{eq:error_system_perturbed},
constrained by $\vect{e}_i \in \mathcal{E}_{i}$, satisfies
$\lim\limits_{t \to \infty} \|\vect{e}_i(t)\| \to 0$, while all
system signals remain bounded in their respective regions as well.

\subsection{Decentralized Control Design} \label{sec:control_design}

Due to the fact that we have to deal with the minimization of norms
$\|\vect{e}_i(t) \|$, as $t \to \infty$, subject to constraints $e_i \in \mathcal{E}_i$,
we invoke here a class of decentralized Nonlinear Model Predictive controllers
(NMPC). NMPC frameworks have been studied in \cite{Mayne2000789, morrari_npmpc, frank_1998_quasi_infinite, cannon_2001_nmpc, camacho_2007_nmpc, fontes_2001_nmpc_stability, frank_2003_towards_sampled-data-nmpc, borrelli_2013_nmpc, grune2016nonlinear} and they have
been proven to be a powerful tool for dealing with state and input constraints. 

Consider a sequence of sampling times $\{t_k\}_{k \in \mathbb{N}}$, with a constant
sampling time $h$, $0 < h < T_p$, where $T_p$ is the finite time predicted horizon,
such that $t_{k+1} = t_k + h$, $\forall k \in \mathbb{N}$. Hereafter we will
denote by $i$ the agent and by index $k$ the sampling instant. In sampled data NMPC, a Finite-Horizon Open-loop Optimal Control Problem (FHOCP) is solved at
discrete sampling time instants $t_k$ based on the current state error measurement
$\vect{e}_i(t_k)$. The solution is an optimal control signal
$\overline{\vect{u}}_i^{\star}(s)$, computed over $s \in [t_k, t_k+T_p]$. The
open-loop input signal applied in between the sampling instants is given by
the solution of the following FHOCP:
\begin{subequations}
	\begin{align}
	&\hspace{-4mm}\min\limits_{\overline{\vect{u}}_i(\cdot)} J_i(\vect{e}_i(t_k), \overline{\vect{u}}_i(\cdot)) \notag \\
	&\hspace{-4mm}= \min\limits_{\overline{\vect{u}}_i(\cdot)} \left\{  V_i(\overline{\vect{e}}_i(t_k+T_p)) + \int_{t_k}^{t_k+T_p} \Big[ F_i(\overline{\vect{e}}_i(s), \overline{\vect{u}}_i(s)) \Big] ds \right\}  \label{mpc_position_based_cost_2} \\
	&\hspace{-4mm}\text{subject to:} \notag \\
	&\hspace{1mm} \dot{\overline{\vect{e}}}(s) = g_i(\overline{\vect{e}}_i(s), \overline{\vect{u}}_i(s)), \overline{\vect{e}}_i(t_k) = \vect{e}_i(t_k), \label{eq:diff_mpc} \\
	&\hspace{1mm} \overline{\vect{e}}_i (s) \in \mathcal{E}_{i, s - t_k}, \overline{\vect{u}}_i(s) \in \mathcal{U}_i, s \in [t_k,t_k+T_p],  \label{eq:mpc_constrained_set}\\
	&\hspace{1mm} \overline{\vect{e}}(t_k+T_p)\in \Omega_i. \label{eq:mpc_terminal_set}
	\end{align}
\end{subequations}
At a generic time $t_k$ then, agent $i \in \mathcal{V}$ solves the aforementioned
FHOCP. The notation $\overline{\cdot}$ is used to distinguish predicted states
which are internal to the controller, corresponding to the nominal system
\eqref{eq:diff_mpc} (i.e., the system \eqref{eq:error_system_perturbed} by
substituting $\vect{w} = \mat{0}_{12 \times 1}$). This means that
$\overline{\vect{e}}_i(\cdot)$ is the solution to \eqref{eq:diff_mpc} driven by
the control input $\overline{\vect{u}}_i(\cdot) : [t_k, t_k + T_p] \to \mathcal{U}_i$
with initial condition $\vect{e}_i(t_k)$. Note that the predicted states are not
the same with the actual closed-loop values due to the fact that the system is
under the presence of disturbances $w_i \in \mathcal{W}_i$, where
$\mathcal{W}_i$ is defined in \eqref{eq:mathcal_WU}. The functions
$F_i : \mathcal{E}_{i} \times \mathcal{U}_i \to \mathbb{R}_{\geq 0}$, $V_i: \mathcal{E}_i \to \mathbb{R}_{\geq 0}$
stand for the \emph{running costs} and the \emph{terminal penalty costs},
respectively, and they are defined by:
\begin{subequations}
	\begin{align}
	F_i \big(\overline{\vect{e}}_i, \overline{\vect{u}}_i\big)
	&\triangleq \overline{\vect{e}}_i^{\top} \mat{Q}_i \overline{\vect{e}}_i
	+ \overline{\vect{u}}_i^{\top} \mat{R}_i \overline{\vect{u}}_i \label{eq:F_i_def}, \\
	V_i \big(\overline{\vect{e}}_i\big) & \triangleq \overline{\vect{e}}_i^{\top} \mat{P}_i \overline{\vect{e}}_i. \label{eq:V_i_def}
	\end{align}
\end{subequations}
$\mat{R}_i \in \mathbb{R}^{6 \times 6}$ and
$\mat{Q}_i, \mat{P}_i \in \mathbb{R}^{12 \times 12}$ are symmetric and positive
definite controller gain matrices to be appropriately tuned. The sets $\mathcal{E}_{i, s - t_k}$, $\Omega_i$ will be explained later.
For the running cost functions $F_i$, $i \in \mathcal{V}$ the following hold:
\begin{lemma} \label{lemma:F_i_bounded_K_class}
	Let the running costs $F_i$ be defined by \eqref{eq:F_i_def}. Then, for all $\vect{\eta}_i \in \mathcal{E}_{i} \times \mathcal{U}_i$, there exist functions $\alpha_1, \alpha_2 \in \mathcal{K}_{\infty}$ such that: $\alpha_1\big(\|\vect{\eta}_i\|\big) \leq F_i\big(\vect{e}_i, \vect{u}_i\big) \leq \alpha_2\big(\| \vect{\eta}_i \|\big), i \in \mathcal{V}$, where $\vect{\eta}_i \triangleq \left[ \vect{e}_i^\top, \vect{u}_i^\top\right]^\top.$
\end{lemma}
\begin{proof}
	The proof can be found in Appendix \ref{app:proof_lemma1}.
\end{proof}
\begin{lemma}\label{lemma:F_Lipschitz}
	The running costs $F_i$ are \emph{locally Lipschitz continuous in} $\mathcal{E}_{i} \times \mathcal{U}_i$. Thus, it holds that: $\big|F_i(\vect{e}_i, \vect{u}_i) - F_i(\vect{e}_i', \vect{u}_i)\big| \leq L_{F_i} \|\vect{e}_i - \vect{e}_i'\|, \forall \vect{e}_i, \vect{e}_i' \in \mathcal{E}_i, \vect{u} \in \mathcal{U}_i,$
	where: $L_{F_i} \triangleq 2 \sigma_{\max}(\mat{Q}_i) \sup\limits_{\vect{e}_i \in \mathcal{E}_{i}} \|\vect{e}_i\|$.
\end{lemma}
\begin{proof}
	The proof can be found in Appendix \ref{app:proof_lemma2}.
\end{proof}

The applied input signal is a portion of the optimal solution to an
optimization problem where information on the states of the neighboring
agents of agent $i$ is taken into account only in the constraints considered
in the optimization problem. These constraints pertain to the set of its
neighbors $\mathcal{N}_i$ and, in total, to the set of all agents within its
sensing range $\mathcal{R}_i$. Regarding these, we make the following assumption:
\begin{assumption} (\textit{Access to Predicted Information from each agent})
	\label{ass:access_to_predicted_info_n}
	When at time $t_k$ agent $i$ solves a FHOCP, it has access to the
	following measurements, across the entire horizon $s \in (t_k, t_k + T_p]$:
	\begin{enumerate}
		\item Measurements of the states:
		\begin{itemize}
			\item $\vect{z}_j(t_k)$ of all agents $j \in \mathcal{R}_i(t_k)$ within
			its sensing range at time $t_k$;
			\item $\vect{z}_{j'}(t_k)$ of all of its neighboring agents $j' \in \mathcal{N}_i$ at time $t_k$;
		\end{itemize}
		\item The \textit{predicted states}:
		\begin{itemize}
			\item $\overline{\vect{z}}_j(s)$ of all agents $j \in \mathcal{R}_i(t_k)$ within its sensing range;
			\item $\overline{\vect{z}}_{j'}(s)$ of all of its neighboring agents $j' \in \mathcal{N}_i$;
		\end{itemize}
	\end{enumerate}
\end{assumption}
\begin{remark}
	The justification for this assumption is as follows. By considering that
	$\mathcal{N}_i \subseteq \mathcal{R}_i(t)$, $\forall t \in \mathbb{R}_{\ge 0}$,
	that the state
	vectors $\vect{z}_j$ are comprised of 12 real numbers encoded by
	4 bytes, and that the sampling occurs with a frequency $f$ for all agents, the
	overall downstream bandwidth required by each agent is:
	$BW_d = 12 \times 32\ \text{[bits]} \times |\mathcal{R}_i| \times \dfrac{T_p}{h} \times f\ [\text{sec}^{-1}]$.
	Given a conservative sampling time $f = 100$ Hz and a horizon of
	$\dfrac{T_p}{h} = 100$ time steps, the wireless protocol IEEE 802.11n-2009
	(a standard for present-day devices) can accommodate up to $|\mathcal{R}_i| = \dfrac{600\ [\text{Mbit}\cdot \text{sec}^{-1}] }{12\times32[\text{bit}]\times10^4 [\text{sec}^{-1}]} \approx
	16 \cdot 10^2 \text{ agents},$ within the range of one agent.
	We deem this number to be large enough for practical applications
	for the approach of assuming access to the predicted states of agents
	within the range of one agent to be reasonable.
\end{remark}

\begin{figure}[t!]
\centering
\begin{tikzpicture}[scale = 0.5]
\draw[dashed] (0,0) -- (10,0);

\node at (-2, 2) {Agent $i$};
\node at (-2, -2) {Agent $j$};

\filldraw[fill=blue!10!white, draw=black](2,2) circle (1cm);
\filldraw[fill=blue!10!white, draw=black](2,-2) circle (1cm);
\node at (2, -4) {$t_k$};

\filldraw[fill=yellow!10!white, draw=black,dashed] (5,2) circle (1cm);
\filldraw[fill=yellow!10!white, draw=black,dashed] (5,-2) circle (1cm);
\node at (5, -4) {$t_{k+1}$};

\filldraw[fill=green!10!white, draw=black,dashed] (8.5, 2.5) circle (1cm);
\filldraw[fill=green!10!white, draw=black,dashed] (8,-2) circle (1cm);
\node at (8, -4) {$t_{k+2}$};

\node at (12, 0) {$\dots$};

\draw[dashed] (14,0) -- (19,0);

\filldraw[fill=red!10!white, draw=black,dashed] (16, 1.5) circle (1cm);
\filldraw[fill=red!10!white, draw=black,dashed] (17,-1.5) circle (1cm);
\node at (16.5, -4) {$t_k + T_p$};
\end{tikzpicture}
	\caption{The inter-agent constraint regime for two agents, $i$, $j$. Fully
		outlined circles denote measured configurations, while partly outlined
		circles denote predicted configurations. During the solution to the
		individual optimization problems, the predicted configuration of each agent
		at each time step is constrained by the predicted configuration of the other
		agent at the same time step (hence the homologously identical colors at
		each discrete time step).}
	\label{fig:constraint_regime_horizon}
\end{figure}
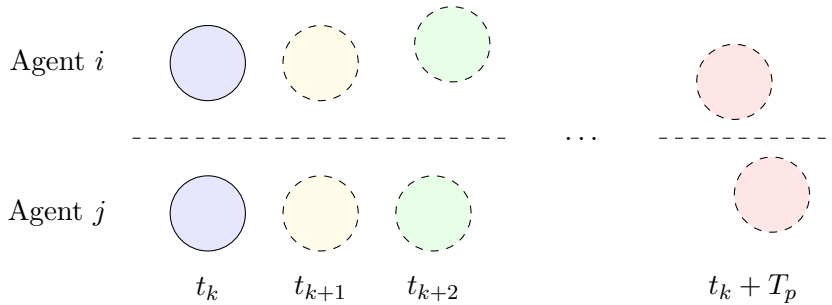

In other words, each time an agent solves its own individual optimization
problem, it knows the (open-loop) state predictions that have been generated
by the solution of the optimization problem of all agents within its sensing
range at that time, for the next $T_p$ time units.
These pieces of information are required, as each agent's trajectory is
constrained not by constant values, but by the trajectories of its associated
agents through time: at each solution time $t_k$ and within the next $T_p$ time
units, an agent's predicted configuration at time $s \in [t_k, t_k + T_p]$ needs
to be constrained by the predicted configuration of its neighboring and
perceivable agents (agents within its sensing range) at the same time
instant $s$, so that collisions are avoided, and connectivity
between neighboring agents is maintained. We assume that the above pieces of
information are \emph{always available}, \emph{accurate} and can be exchanged
\emph{without delay}. Figure \ref{fig:constraint_regime_horizon} depicts the
designed inter-agent (and intra-horizon) constraint regime.

\begin{figure}[t!]\centering
\tikzstyle{decision} = [diamond, draw, 
text width=7.5em, text badly centered, node distance=3cm, inner sep=0pt]
\tikzstyle{block} = [rectangle, draw, 
text width=10em, text centered, minimum height=4em]
\tikzstyle{block_rounded} = [rectangle, rounded corners, draw, 
text width=12em, text centered, minimum height=4em]
\tikzstyle{line} = [draw, -latex']
\tikzstyle{cloud} = [draw, ellipse, 
node distance=3cm, minimum height=4em, minimum width=4em]

\scalebox{0.6}{\begin{tikzpicture}[node distance = 2cm, auto]


\node [block, node distance=1.5cm] (system_m) {Agent $m \in \mathcal{R}_i(t_k)$};
\node [block, below of=system_m, node distance=2.5cm] (system_n) {Agent $n \in \mathcal{R}_i(t_k)$};

\node[right of=system_n, node distance=4.5cm] (right_of_n){};

\node [block_rounded, below of=right_of_n, node distance=2.5cm] (latest_plans) {Latest predictions (current timestep $t_k$)};
\node [block, below of=latest_plans, node distance=2.5cm] (agent_i) {Agent $i$};
\node [block_rounded, below of=agent_i, node distance=2.5cm] (latest_plans_) {Latest predictions \\ (previous timestep $t_{k-1}$)};

\node[left of=latest_plans_, node distance=4.5cm] (left_of_i){};

\node [block, below of=left_of_i, node distance=2.5cm] (system_p) {Agent $p \in \mathcal{R}_i(t_k)$};
\node [block, below of=system_p, node distance=2.5cm] (system_q) {Agent $q \in \mathcal{R}_i(t_k)$};

\node [above of=system_m, node distance=1.5cm] (dots_0) {$\vdots$};
\node [below of=system_q, node distance=1.5cm] (dots_1) {$\vdots$};

\node[inner sep=0,minimum size=0,right of=system_m, node distance=5.5cm] (sm) {};
\node[inner sep=0,minimum size=0,right of=system_n, node distance=4.5cm] (sn) {};
\node[inner sep=0,minimum size=0,right of=system_p, node distance=4.5cm] (sp) {};
\node[inner sep=0,minimum size=0,right of=system_q, node distance=5.5cm] (sq) {};

\path[line]           (system_m) -- (sm);
\path[line]           (system_n) -- (sn);
\path[line]           (sm) -- (latest_plans.38);
\path[line]           (sn) -- (latest_plans);

\path[line]           (system_p) -- (sp);
\path[line]           (system_q) -- (sq);
\path[line]           (sp) -- (latest_plans_);
\path[line]           (sq) -- (latest_plans_.-38);

\path[line,dashed]    (latest_plans) -- (agent_i);
\path[line,dashed]    (latest_plans.-38) -- (agent_i.38);

\path[line,dashed]    (latest_plans_) -- (agent_i);
\path[line,dashed]    (latest_plans_.38) -- (agent_i.-38);
\end{tikzpicture}}
\caption{The flow of information to agent $i$ regarding his perception of
		agents within its sensing range $\mathcal{R}_i$ at arbitrary FHOCP
		solution time $t_k$. Agents $m,n \in \mathcal{R}_i(t_k)$ have solved their
		FHOCP; agent $i$ is next; agents $p,q \in \mathcal{R}_i(t_k)$ have not
		solved their FHOCP yet.}
\label{fig:information_flow}
\end{figure}
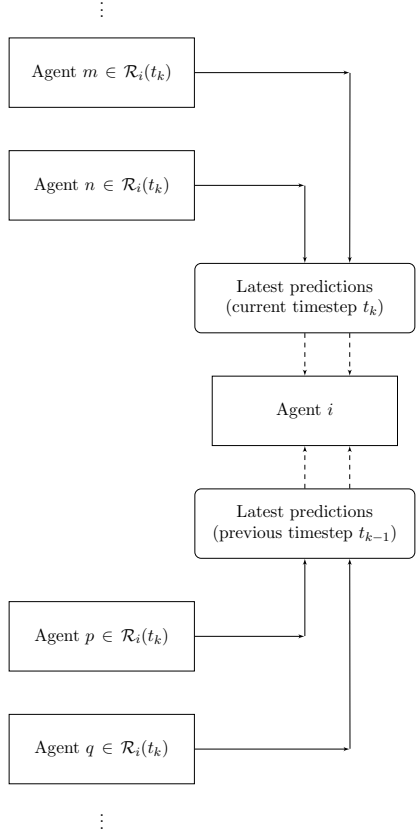

\begin{remark}
	The designed procedure flow can be either concurrent or sequential,
	meaning that agents can solve their individual FHOCP’s and apply the control
	inputs either simultaneously, or one after the other. The conceptual
	design itself is procedure-flow agnostic, and hence it can incorporate both
	without loss of feasibility or successful stabilization. The approach that we
	have adopted here is the sequential one: each agent solves its own FHOCP and
	applies the corresponding admissible control input in a round robin way,
	considering the current and planned (open-loop state predictions)
	configurations of all agents within its sensing range.
	This choice is made on account of three reasons:
		(a) Safety: if a parallel approach is adopted, at the limit, that is in the
		event that the communication (maximum distance) range is comparable to the
		size of the agents (consider for instance the case where two UAVs need to
		collaboratively transport a similar-sized object) it is more likely for
		collisions to occur. This is because, within a parallel approach, agents
		would need to rely more on the open-loop predictions of their neighboring
		agents, which, in the case of disturbances, would make the violation of
		constraints more likely to occur.
		(b) Conversely, the sequential approach allows each agent, in turn, to have access to
		the direct measurements of all other agents' position and overall configuration,
		as well as their predicted trajectories, \textit{before} solving its own optimization
		problem, thereby allowing agents to plan their trajectories and execute their
		motions using additional and concrete information on top of the (in principle
		approximate) open-loop predictions, and therefore without danger of violating
		their constraints. (c) Synchronization problems between agents are avoided within a sequential
		approach, since, from the moment when agent $i \in \mathcal{V}$ starts to solve
		its optimization problem to the moment that it concludes executing its motion,
		all agents $j \in \mathcal{V}, j \neq i$ are assumed stationary, while the
		configuration of those within its sensing range is known to $i$. Figure \ref{fig:process_flow} and Figure \ref{fig:information_flow} depict
	the sequential procedural and informational regimes.
\end{remark}

\begin{figure}[t!]\centering
\tikzstyle{decision} = [diamond, draw, 
text width=7.5em, text badly centered, node distance=3cm, inner sep=0pt]
\tikzstyle{block} = [rectangle, draw, 
text width=7em, text centered, minimum height=4em]
\tikzstyle{line} = [draw, -latex']
\tikzstyle{cloud} = [draw, ellipse, 
node distance=3cm, minimum height=4em, minimum width=4em]

\scalebox{0.6}{\begin{tikzpicture}[node distance = 2cm, auto]
\node [decision] (decide) {Current configuration feasible?};
\node [cloud, below of=decide, node distance=5.5cm] (stop) {stop};

\node [block, right of=decide, node distance=5cm] (solve_1) {Agent 1: solve FHOCP and apply control input};
\node [block, below of=solve_1, node distance=2.5cm] (solve_2) {Agent 2: solve FHOCP and apply control input};
\node [block, below of=solve_2, node distance=3cm] (solve_N) {Agent N: solve FHOCP and apply control input};

\node [block, right of=solve_1, node distance=5cm] (solve_1_) {Agent 1: solve FHOCP and apply control input};
\node [block, below of=solve_1_, node distance=2.5cm] (solve_2_) {Agent 2: solve FHOCP and apply control input};
\node [block, below of=solve_2_, node distance=3cm] (solve_N_) {Agent N: solve FHOCP and apply control input};

\node [above of=decide, node distance=3cm] (step_0) {STEP 0};
\node [above of=solve_1, node distance=3cm] (step_1) {STEP 1};
\node [above of=solve_1_, node distance=3cm] (step_1_) {STEP 2};

\node[inner sep=0,minimum size=0,right of=solve_N, node distance=2.5cm] (sN) {};
\node[inner sep=0,minimum size=0,left of=solve_1_, node distance=2.5cm] (s1_) {};
\node[inner sep=0,minimum size=0,right of=solve_N_, node distance=2.5cm] (sN_) {};
\node[inner sep=0,minimum size=0,right of=solve_1_, node distance=2.5cm] (s1_inv) {};
\node[right of=solve_1_, node distance=3.5cm] (solve_1_inv) {$\dots$};

\path [line]        (decide)  -- node {NO}(stop);
\path [line]        (decide)  -- node {YES}(solve_1);
\path [line]        (solve_1) -- node {}(solve_2);
\path [line,dashed] (solve_2) -- node {}(solve_N);
\path [line]        (solve_1_) -- node {}(solve_2_);
\path [line,dashed] (solve_2_) -- node {}(solve_N_);

\path [line] (solve_N) -- (sN);
\path [line] (sN) -- (s1_);
\path [line] (s1_) -- (solve_1_);

\path [line] (solve_N_) -- (sN_);
\path [line] (sN_) -- (s1_inv);
\path [line] (s1_inv) -- (solve_1_inv);
\end{tikzpicture}}
	\caption{The procedure is approached sequentially. Notice that the
		figure implies that recursive feasibility is established if the initial
		configuration is itself feasible.}
	\label{fig:process_flow}
\end{figure}
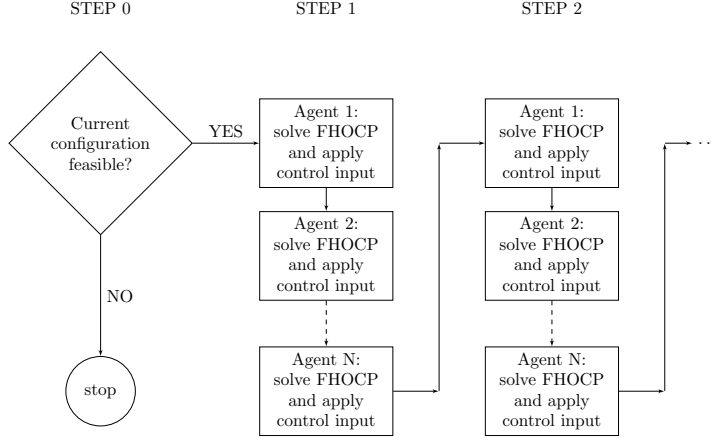

The solution to FHOCP \eqref{mpc_position_based_cost_2} - \eqref{eq:mpc_terminal_set}
at time $t_k$ provides an optimal control input, denoted by
$\overline{\vect{u}}_i^{\star}(s;\ \vect{e}_i(t_k))$, $s \in [t_k, t_k + T_p]$. This
control input is then applied to the system until the next sampling instant $t_{k+1}$:
\begin{align}
\vect{u}_i(s; \vect{e}_i(t_k)) = \overline{\vect{u}}_i^{\star}\big(s; \ \vect{e}_i(t_k)\big),\  s \in [t_k, t_{k+1}).
\label{eq:position_based_optimal_u_2}
\end{align}
At time $t_{k+1}$ a new finite horizon optimal control problem is solved in the
same manner, leading to a receding horizon approach.

The control input $\vect{u}_i(\cdot)$ is of feedback form, since it is recalculated
at each sampling instant based on the then-current state. The solution of
\eqref{eq:error_system_perturbed} at time $s$, $s \in [t_k, t_k+T_p]$, starting
at time $t_k$, from an initial condition $\vect{e}_i(t_k) = \overline{\vect{e}}_i(t_k)$,
by application of the control input $\vect{u}_i : [t_k, s] \to \mathcal{U}_i$ is
denoted by: $\vect{e}_i\big(s;\ \vect{u}_i(\cdot), \vect{e}_i(t_k)\big),\ s \in [t_k, t_k+T_p]$.

The \textit{predicted} state of the system \eqref{eq:diff_mpc} at time $s$,
$s \in [t_k, t_k+T_p]$ based on the measurement of the state at time $t_k$,
$\vect{e}_i(t_k)$, by application of the control input
$\vect{u}_i\big(s;\ \vect{e}_i(t_k)\big)$, for the time period $s \in [t_k, t_k + T_p]$ is denoted by: $\overline{\vect{e}}_i\big(s;\ \vect{u}_i(\cdot), \vect{e}_i(t_k)\big), s \in [t_k, t_k+T_p]$.

Due to the fact that the system is in presence of disturbances
$w_i \in \mathcal{W}_i$, as $\mathcal{W}_i$ defined in \eqref{eq:mathcal_WU}, it holds in general that: $\overline{\vect{e}}_i(\cdot) \neq \vect{e}_i(\cdot)$.

\begin{property}
	By integrating \eqref{eq:error_system_perturbed}, \eqref{eq:diff_mpc} at the
	time interval $s \ge \tau$, the actual $\vect{e}_i(\cdot)$ and the predicted
	states $\overline{\vect{e}}_i(\cdot)$ are respectively given by:
	\begin{subequations}
		\begin{align}
		\vect{e}_i\big(s;\ \vect{u}_i(\cdot), \vect{e}_i(\tau)\big) &=
		\vect{e}_i(\tau) + \int_{\tau}^{s} h_i\big(\vect{e}_i(s';\ \vect{e}_i(\tau)), \vect{u}_i(s)\big) ds', \label{eq:remark_4_eq_1} \\
		\overline{\vect{e}}_i\big(s;\ \vect{u}_i(\cdot), \vect{e}_i(\tau)\big) &=
		\vect{e}_i(\tau) + \int_{\tau}^{s} g_i\big(\overline{\vect{e}}_i(s';\ \vect{e}_i(\tau)), \vect{u}_i(s')\big) ds'. \label{eq:remark_4_eq_2}
		\end{align}
	\end{subequations}
	\label{remark:predicted_actual_equations_with_disturbance}
\end{property}
The satisfaction of the constraints $\mathcal{E}_i$ on the state along the
prediction horizon depends on the future realization of the uncertainties. Through the assumption of additive uncertainty and Lipschitz continuity
of the nominal model, it is possible to compute a bound on the future effect of
the uncertainty on the system. Then, by considering this effect on the state
constraint on the nominal prediction, it is possible to guarantee that the
evolution of the real state of the system will be admissible for all times.
In view of the latter, the state constraint set $\mathcal{E}_i$ of the standard
NMPC formulation, is being replaced by a restricted constraint set
$\mathcal{E}_{s-t_k} \subseteq \mathcal{E}_i$ in \eqref{eq:mpc_constrained_set}.
This state constraints' tightening for the nominal system \eqref{eq:diff_mpc}
with additive disturbance $\vect{w}_i \in \mathcal{W}_i$ is a key ingredient of
the proposed controller and guarantees that the evolution of the evolution of
the real system will be admissible for all times. If the state constraint set
was left unchanged during the solution of the optimization problem, the applied
input to the plant, coupled with the uncertainty affecting the states of the
plant could force the states of the plant to escape their intended bounds. The aforementioned tightening set strategy is inspired by the works
\cite{1185106, Fontes2007, alina_ecc_2011}.

\begin{lemma} \label{lemma:diff_state_from_same_conditions}
	The difference between the actual measurement $\vect{e}_i\big(t_k + s;\ \vect{u}_i(\cdot), \vect{e}_i(t_k)\big)$ at time $t_k+s$, $s \in (0, T_p]$, and the predicted state $\overline{\vect{e}}_i\big(t_k + s;\ \vect{u}_i(\cdot), \vect{e}_i(t_k)\big)$ at the same time, under a control input $\vect{u}_i(\cdot) \in \mathcal{U}_i$, starting at the same initial state $\vect{e}_i(t_k)$ is upper bounded by: $\left\| \vect{e}_i\big(t_k + s;\vect{u}_i(\cdot), \vect{e}_i(t_k)\big) -
	\overline{\vect{e}}_i\big(t_k + s;\vect{u}_i(\cdot), \vect{e}_i(t_k)\big) \right\| \leq \dfrac{\overline{w}_i}{L_{g_i}} (e^{L_{g_i} s} - 1)$, $s \in (0, T_p]$, where $\overline{w}_i$ is the upper bound of the disturbance as defined in \eqref{eq:mathcal_WU}, and $L_{g_i}$ is defined in \eqref{eq:lipsitch_g_i}.
\end{lemma}
\begin{proof}
	The proof can be found in Appendix \ref{proof:lemma_diff_state_from_same_conditions}.
\end{proof}
\noindent By taking into consideration the aforementioned Lemma, the restricted constraints set are then defined by: $\mathcal{E}_{i, s-t_k} \triangleq \mathcal{E}_i \ominus \mathcal{X}_{i,s-t_k}$, where:
\begin{align}
\mathcal{X}_{i,s-t_k} = \left\{ \vect{e}_i \in \mathcal{M} \times \mathbb{R}^6 :
\|\vect{e}_i(s)\| \leq \dfrac{\overline{w}_i}{L_{g_i}}\big( e^{L_{g_i}(s - t_k)} - 1\big),\ \forall s \in [t_k, t_k + T_p] \right\}.
\label{eq:b_restricted_constraint_set}
\end{align}
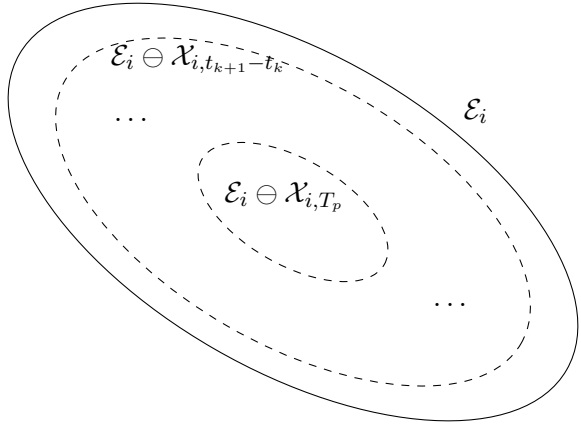
\begin{figure}[t!]
	\centering
\begin{tikzpicture}[scale = 0.7, rotate=-30]
\draw (2,2) ellipse (6cm and 3cm);
\node at ($(2.5,2.5)+(75:6 and 3)$) {$\mathcal{E}_i$};
\draw[dashed] (2,2) ellipse (5cm and 2.5cm);
\node at ($(-2.3,1.2)+(75:5 and 2.5)$) {$\mathcal{E}_i \ominus \mathcal{X}_{i,t_{k+1} - t_k}$};
\draw[dashed] (2,2) ellipse (2cm and 1cm);
\node at ($(1.2,1.2)+(75:2 and 1)$) {$\mathcal{E}_i \ominus \mathcal{X}_{i, T_p}$};

\node at (5.5,2) {$\dots$};
\node at (-1.5,2) {$\dots$};

\pgflowlevel{\pgftransformrotate{30}}
\end{tikzpicture}
	\caption{The nominal constraint set $\mathcal{E}_i$ in bold, and the
		consecutive restricted constraint sets $\mathcal{E}_i \ominus \mathcal{X}_{i, s-t_k}$,
		$s \in [t_k, t_k + T_p]$, dashed. The predicted state is constrained
		by a different and more tight set at each different time instant, since the
		more times goes by the more uncertain the true state becomes, and hence, the
		more the model state needs to be restricted if the true state is to be
		constrained within $\mathcal{E}_i$ at all times.}
	\label{fig:tightening_high_level}
\end{figure}

If the state constraint set considered in the solution of the FHOCP is given by: $\mathcal{E}_{i, s-t_k}$, then the state of the real system $\vect{e}_i$ is guaranteed to fulfill the original state constraint sets $\mathcal{E}_i$. We formalize this statement in Property \ref{property:restricted_constraint_set}.
\begin{property}
	\label{property:restricted_constraint_set}
	For every $s \in [t_k, t_k + T_p]$, it holds that if: $\overline{\vect{e}}_i\big( s;\ \vect{u}_i(\cdot,\ \vect{e}_i(t_k)), \vect{e}_i(t_k) \big) \in \mathcal{E}_i \ominus \mathcal{X}_{i,s-t_k}$,
	where $\mathcal{X}_{i,s-t_k}$ is given by \eqref{eq:b_restricted_constraint_set}, then the real state $\vect{e}_i$ satisfies the constraints $\mathcal{E}_i$, i.e., $\vect{e}_i(s) \in \mathcal{E}_i$.
\end{property}
\begin{proof}
	The proof can be found in Appendix \ref{app:property_2}.
\end{proof}
\begin{assumption}
	\label{ass:psi}
	The terminal set $\Omega_i$ is a subset of an admissible and positively
	invariant set $\Psi_i$, with $\Omega_i \subseteq \Psi_i$, where $\Psi_i$ is defined by: $\Psi_i \triangleq \big\{\overline{\vect{e}}_i \in \Phi_i : V_i(\overline{\vect{e}}_i) \leq \varepsilon_{\Psi_i} \big\},\ \varepsilon_{\Psi_i} > 0$.
\end{assumption}
\begin{assumption} \label{ass:psi_psi}
	The set $\Psi_i$ is interior to the set $\Phi_i$, $\Psi_i \subseteq \Phi_i$,
	which is the set of states within $\mathcal{E}_{i,T_p-h}$ for which
	there exists an admissible control input (see Definition
	\ref{definition:admissible_input_with_disturbance}) which is of linear
	feedback form with respect to the state $\kappa_i(\vect{e}_i) : [0,h] \to \mathcal{U}_i$: $\Phi_i \triangleq \big\{\overline{\vect{e}}_i \in \mathcal{E}_{i,T_p-h} : \kappa_i(\overline{\vect{e}}_i) \in \mathcal{U}_i \big\}$, such that for all $\vect{e}_i \in \Psi_i$ and for all $s \in [0,h]$ it holds that:
	\begin{align}
	\dfrac{\partial V_i}{\partial \vect{e}_i} g_i\big(\vect{e}_i(s), \kappa_i(\vect{e}_i(s))\big)
	+ F_i\big(\vect{e}_i(s), \kappa_i(\vect{e}_i(s))\big) \leq 0. \label{eq:phi_psi}
	\end{align}
\end{assumption}

\begin{remark} \label{remark:aux_control_stabilizability}
	The existence of the robust linear state-feedback control law $\kappa_i$ is ensured if:
	1) the linearization of system \eqref{eq:error_system_perturbed} is stabilizable;
	2) the function $h_i$ is twice differentiable, \emph{locally Lipschitz continuous in $\mathcal{E}_i \times \mathcal{U}_i$} with $f(\vect{0},\vect{0}) = \vect{0}$, and
	3) $\mathcal{U}_i$ is a compact subset of $\mathbb{R}^6$ containing the origin in its interior \cite{262032, FINDEISEN2003190}.
\end{remark}
\begin{assumption}
	\label{ass:psi_omega}
	The admissible and positively invariant set $\Psi_i$ is such that $\forall \vect{e}_i(t) \in \Psi_i \Rightarrow \overline{\vect{e}}_i\big(t+s;\ \kappa_i(\vect{e}_i(t)), \vect{e}_i(t)\big) \in \Omega_i \subseteq \Psi_i$,
	for some $s \in [0,h]$.
\end{assumption}
\noindent The terminal sets $\Omega_i$ are chosen to be closed, including the origin, as: $\Omega_i \triangleq \big\{\overline{\vect{e}}_i \in \mathcal{E}_i : V_i(\overline{\vect{e}}_i)$ $\leq \varepsilon_{\Omega_i}\big\}\ \text{, where } \varepsilon_{\Omega_i} \in (0, \varepsilon_{\Psi_i})$.
\begin{remark}
	It should be noted that the larger the length of the time-horizon $T_p$ the
	more probable (in general) it becomes that the sets $\mathcal{E}_{i,s}$ may
	become empty beyond some $s \in [t_k, t_k + T_p]$. The length of the
	time-horizon should hence be designed so that the above violation does not
	occur.  \label{remark:E_T_P_limit}
\end{remark}
\noindent For the terminal cost penalty functions $V_i$, $i \in \mathcal{V}$ the following hold:
\begin{lemma} \label{lemma:V_i_lower_upper_bounded}
	Let the functions $V_i$ be defined by \eqref{eq:V_i_def}. Then, for every $\vect{e}_i \in \Psi_i$ there exist functions $\alpha_1, \alpha_2 \in \mathcal{K}_{\infty}$ such that: $\alpha_1\big(\|\vect{e}_i\|\big) \leq V_i(\vect{e}_i) \leq \alpha_2\big(\| \vect{e}_i \|\big)$, $\forall i \in \mathcal{V}$.
\end{lemma}
\begin{proof}
	The proof can be found in Appendix \ref{appendix_lemma_4}.
\end{proof}
\begin{lemma} \label{lemma:V_Lipschitz_e_0}
	The terminal penalty functions $V_i$ are locally Lipschitz continuous in $\Psi_i$. Thus it holds that: $\big|V_i(\vect{e}_i) - V_i(\vect{e}_i')\big| \leq L_{V_i} \|\vect{e}_i - \vect{e}_i'\|$, $\forall \vect{e}_i, \vect{e}_i' \in \Psi_i,$ where: $L_{V_i} = 2 \sigma_{\max}(\mat{P}_i) \sup \limits_{\vect{e}_i \in \Psi_i} \|\vect{e}_i\|$.
\end{lemma}
\begin{proof}
	The proof is similar to the proof of Lemma \ref{lemma:F_Lipschitz} and is omitted.
\end{proof}
\begin{figure}[t!]
	\centering
\begin{tikzpicture}[scale = 0.7, rotate=-30]
\draw[dashed](2,2) ellipse (5cm and 2.5cm);
\node at ($(2.7,2.7)+(75:5 and 2.5)$) {$\mathcal{E}_i \ominus \mathcal{X}_{T_p-h}$};
\draw[dashdotted](2,2) ellipse (4cm and 2cm);
\node at ($(2.2,2.2)+(75:4 and 2)$) {$\Phi_i$};
\draw[dashdotted] (2,2) ellipse (3cm and 1.5cm);
\node at ($(2.2,2.2)+(75:3 and 1.5)$) {$\Psi_i$};
\draw (2,2) ellipse (2cm and 1cm);
\node at ($(2.2,2.2)+(75:2 and 1)$) {$\Omega_i$};
\pgflowlevel{\pgftransformrotate{30}}
\end{tikzpicture}
	\caption{The hierarchy of sets
		$\Omega_i \subseteq \Psi_i \subseteq \Phi_i \subseteq \mathcal{E}_{i,T_p-h}$,
		in bold, dash-dotted, dash-dotted, and dashed, respectively.
		For every state in $\Phi_i$ there is a linear state feedback control
		$\kappa_i(\vect{e}_i)$ which, when applied to a state
		$\vect{e}_i \in \Psi_i$, forces the trajectory of the state of the system to reach the terminal set $\Omega_i$.}
	\label{fig:tightening_low_level}
\end{figure}
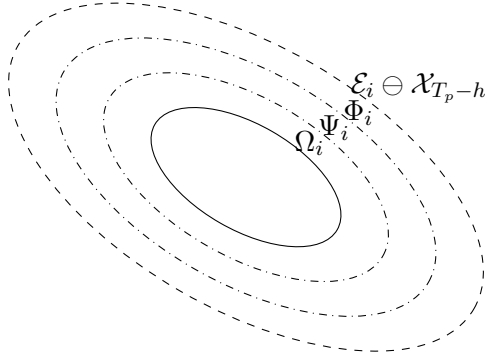
We can now give the definition of an \textit{admissible input} for the FHOCP \eqref{mpc_position_based_cost_2}-\eqref{eq:mpc_terminal_set}.
\begin{definition} (Admissible input for FHOCP \eqref{mpc_position_based_cost_2}-\eqref{eq:mpc_terminal_set})  \label{definition:admissible_input_with_disturbance}
	A control input $\vect{u}_i : [t_k, t_k + T_p] \to \mathbb{R}^6$ for a state
	$\vect{e}_i(t_k)$ is called \textit{admissible} for the problem
	\eqref{mpc_position_based_cost_2}-\eqref{eq:mpc_terminal_set} if the following hold: $1)$ $\vect{u}_i(\cdot)$ is piecewise continuous; $2)$ $\vect{u}_i(s) \in \mathcal{U}_i,\ \forall s \in [t_k, t_k + T_p]$; $3)$ $\overline{\vect{e}}_i\big(t_k + s;\ \vect{u}_i(\cdot), \vect{e}_i(t_k)\big) \in \mathcal{E}_i \ominus \mathcal{X}_{i,s},\ \forall s \in [0, T_p]$; $4)$ $\overline{\vect{e}}_i\big(t_k + T_p;\ \vect{u}_i(\cdot), \vect{e}_i(t_k)\big) \in \Omega_i$;
	
	In other words, $\vect{u}_i$ is admissible if it conforms to the constraints
	on the input and its application yields states that conform to the
	prescribed state constraints of FHOCP \eqref{mpc_position_based_cost_2}-\eqref{eq:mpc_terminal_set}
	along the entire horizon $[t_k, t_k + T_p]$, and the terminal predicted
	state conforms to the terminal constraint.
\end{definition}

Under these considerations, we can now state the theorem that relates to
the guaranteeing of the stability of the compound system of agents
$i \in \mathcal{V}$, when each of them is assigned a desired position and orientation:

\begin{theorem} \label{theorem}
	\label{theorem:with_disturbances}
	Suppose that for every $i \in \mathcal{V}$:
	
	\begin{enumerate}
		\item Assumptions \ref{ass:g_i_g_R_Lipschitz}-\ref{ass:psi_omega} hold;
		\item A solution to FHOCP \eqref{mpc_position_based_cost_2}-\eqref{eq:mpc_terminal_set} is feasible at time $t=0$ with feasible initial conditions, as defined in Definition \ref{def:set_feasible_initial_conditions};
		\item The upper bound $\overline{w}_i$ of the disturbance
		$\vect{w}_i$ satisfies the following:
		\begin{align} \label{eq:upper_bound}
		\overline{w}_i \leq \dfrac{\varepsilon_{\Psi_i} - \varepsilon_{\Omega_i}}{\dfrac{L_{V_i}}{L_{g_i}} (e^{L_{g_i}h} - 1) e^{L_{g_i} (T_p - h)}},
		\end{align}
		for all $t \in \mathbb{R}_{\geq 0}$.
	\end{enumerate}
	Then the closed loop trajectories of the system \eqref{eq:error_system_perturbed}, under the control input \eqref{eq:position_based_optimal_u_2} which is the outcome of the FHOCP \eqref{mpc_position_based_cost_2}-\eqref{eq:mpc_terminal_set}, converge to the set $\Omega_i$, as $t \to \infty$ and are ultimately bounded there, for every $i \in \mathcal{V}$.
\end{theorem}

\begin{proof}
	The proof of the above theorem consists of two parts: in the first, recursive feasibility is established, that is, initial
	feasibility is shown to imply subsequent feasibility; in the second, and based
	on the first part, it is shown that the error state $\vect{e}_i(t)$ reaches
	the terminal set $\Omega_i$ and is trapped there. The feasibility analysis can be found in Appendix \ref{app:feasibility}. The convergence analysis can be found in Appendix \ref{app:convergence}.
\end{proof}

\begin{remark}
	Inequality \eqref{eq:upper_bound} gives an upper bound of the disturbance that the proposed methodology can handle. Disturbances excheeding this bound cannot guarantee the feasibility of Theorem \ref{theorem}.
\end{remark}

\begin{remark}
	Due to the existence of disturbances, the position and orientation error
	of each agent cannot be made to become arbitrarily close to zero, and
	therefore $\lim\limits_{t \to \infty} \|\vect{e}_i(t)\|$ cannot converge to zero.
	However, if the conditions of Theorem 2 hold, then this error can be bounded
	above by the quantity $\sqrt{\varepsilon_{\Omega_i} / \lambda_{\max}(\mat{P}_i)}$
	(since the trajectory of the error is trapped in the terminal set, this means
	that $V(\vect{e}_i) = \vect{e}_i^{\top} \mat{P}_i \vect{e}_i \leq \varepsilon_{\Omega_i}$).
\end{remark}

\begin{remark}
In sampled-data Model Predictive Control, the solution to the optimization problem is the input that is implemented on the continuous time system \eqref{eq:error_system_perturbed}. However, the solution of the FHOCP \eqref{eq:mpc_constrained_set}-\eqref{eq:mpc_terminal_set} is computed in a discrete-time manner. In order to address this, the input is held constant over the time period between successive solutions of the optimization problem using zero-order hold. For  more details we refer the reader to \cite{frank_2003_towards_sampled-data-nmpc}. Implementation tools of this approach, which we also have adopted for our simulation experiments (see next section), can be found in \cite{grune2016nonlinear}.
\end{remark}

\section{Simulation Results}
\label{sec:simulation_results}

For a simulation scenario, consider $N = 3$ unicycle agents with dynamics:
$\dot{\vect{z}}_i(t) =
\begin{bmatrix}
\dot{x}_i(t) \\
\dot{y}_i(t) \\
\dot{\theta}_i(t) \\
\end{bmatrix}
=
\begin{bmatrix}
v_i(t) \cos \theta_i(t) \\
v_i(t) \sin \theta_i(t) \\
\omega_i(t) \\
\end{bmatrix}
+
w_i(t)
\begin{bmatrix}
1 \\
1 \\
1 \\
\end{bmatrix}
$, $i \in \mathcal{V} = \{1,2,3\}$, where: $\vect{z}_i = \left[x_i, y_i, \theta_i \right]^\top$, $f_i(\vect{z}_i, \vect{u}_i) = \left[v_i \cos \theta_i, v_i \sin \theta_i, \omega_i \right]^\top$,
$\vect{u}_i = \left[v_i, \omega_i \right]^\top$, $w_i = \overline{w}_i \sin(2 t)$,
with $\overline{w}_i = 0.1$. For the control inputs we set $\overline{u}_i = 8\sqrt{2}$.
The radius of the agents is $r_i = 0.5$. The sensing range of all agents is
$d_i = 4r_i = 2.0$. Their obstacle-detection range is set to $b_i = 4.0$.
We set $\varepsilon = 0.01$, where $\varepsilon$ is the parameter of the constraint set $\mathcal{Z}_i$. The neighboring sets are set
to $\mathcal{N}_1 = \{2,3\}$, $\mathcal{N}_2 = \mathcal{N}_3 = \{1\}$. Agent $3$ is chosen to execute motions first, then agent $1$, followed by
	agent $2$. The agents' initial positions are $\vect{z}_1$ $=$ $[-6, 3.5, 0]^{\top}$,
$\vect{z}_2$ $=$ $[-6, 2.3, 0]^{\top}$ and $\vect{z}_3$ $=$ $[-6, 4.7, 0]^{\top}$.
Their desired configurations in steady-state are $\vect{z}_{1, \text{des}}$ $=$
$[6, 3.5, 0]^{\top}$, $\vect{z}_{2, \text{des}}$ $=$ $[6, 2.3, 0]^{\top}$ and
$\vect{z}_{3, \text{des}}$ $=$ $[6, 4.7, 0]^{\top}$. In the workspace, we place
$2$ obstacles with centers at points $[0, 2.0]^{\top}$ and $[0, 5.5]^{\top}$,
respectively. The obstacles' radii are $r_{\scriptscriptstyle O_\ell} = 1.0$,
$\ell \in \mathcal{L} = \{1,2\}$. The matrices $\mat{Q}_i$, $\mat{R}_i$,
$\mat{P}_i$ are set to $\mat{Q}_i = 0.5 (I_3 + 0.5\dagger_3)$, $\mat{R}_i = 0.005 [5 \ 0; 0 \ 1]$ and $\mat{P}_i = 0.3 (I_3 + 0.5\dagger_3)$, where $\dagger_N$ is a $N \times N$ matrix whose elements are randomly chosen between the values $0.0$ and $1.0$.
The maximum eigenvalue of matrix $\mat{P}_i$ was found to be $\lambda_{\max}(\mat{P}_i) = 0.4710$.
The sampling time is $h = 0.1$ sec, the time-horizon is $T_p = 0.6$ sec, and the
total execution time given is $10$ sec. Furthermore, we set: $L_{f_i} = 8.5883$,
	$L_{V_i} = 0.0471$, $\varepsilon_{\Psi_i} = 0.0582$ and $\varepsilon_{\Omega_i} = 0.0035$
for all $i \in \mathcal{V}$.

\begin{figure}[t!]
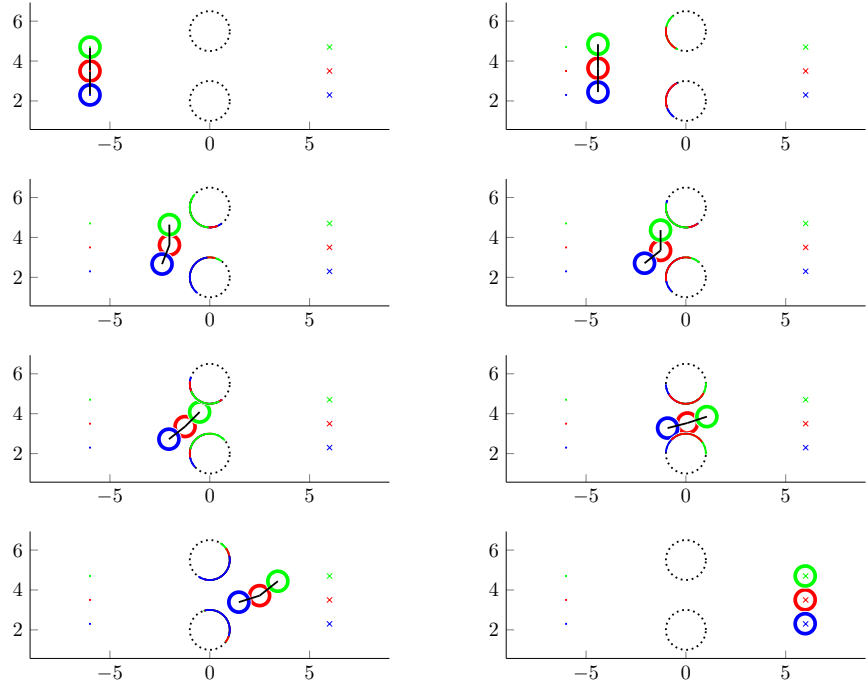
\centering
	\scalebox{0.7}{
}
	\caption{The trajectories of the three agents in the $x-y$ plane. Agent 1 is in
			red, agent 2 in blue and agent 3 in green. Agent $3$ executes its motions first,
				followed by agent $1$ and then agent $2$. A faint black line connects
			agents deemed neighbors. A point on the circumference of the obstacles is
			black and dotted when it is not visible by an agent; otherwise it is colored
			in accordance with which agent it is visible. Mark X marks the desired configurations.}
	\label{fig:d_OFF_res_trajectory_3_2}
\end{figure}

\begin{figure}[t!]
	\centering
	\includegraphics[scale = 0.50]{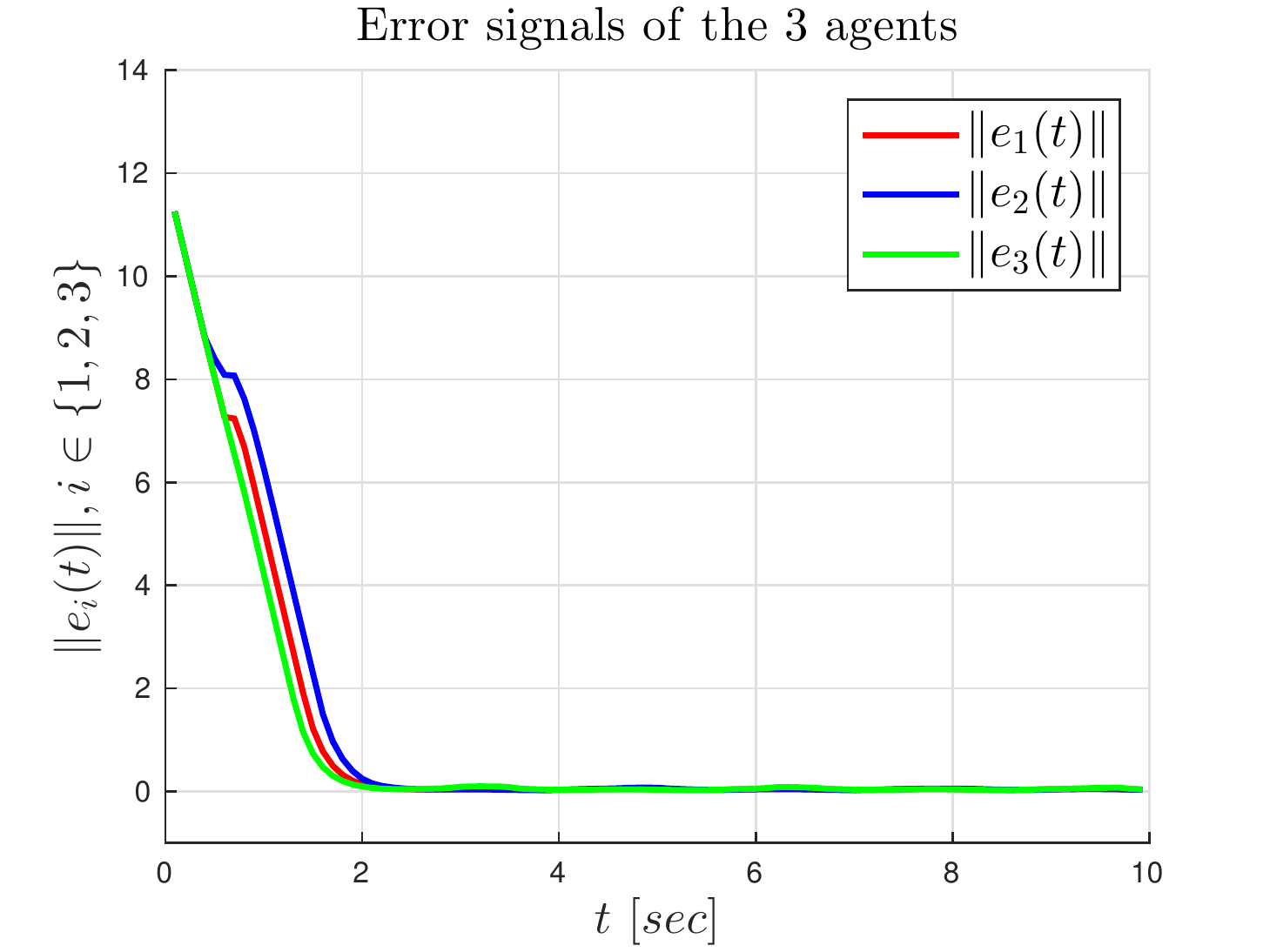}
	\caption{The evolution of the $2-$norms of the error signals of the three
			agents over time.}
	\label{fig:d_ON_res_3_2_errors_agent_1}
\end{figure}

\begin{figure}[t!]
	\centering
	\includegraphics[scale = 0.50]{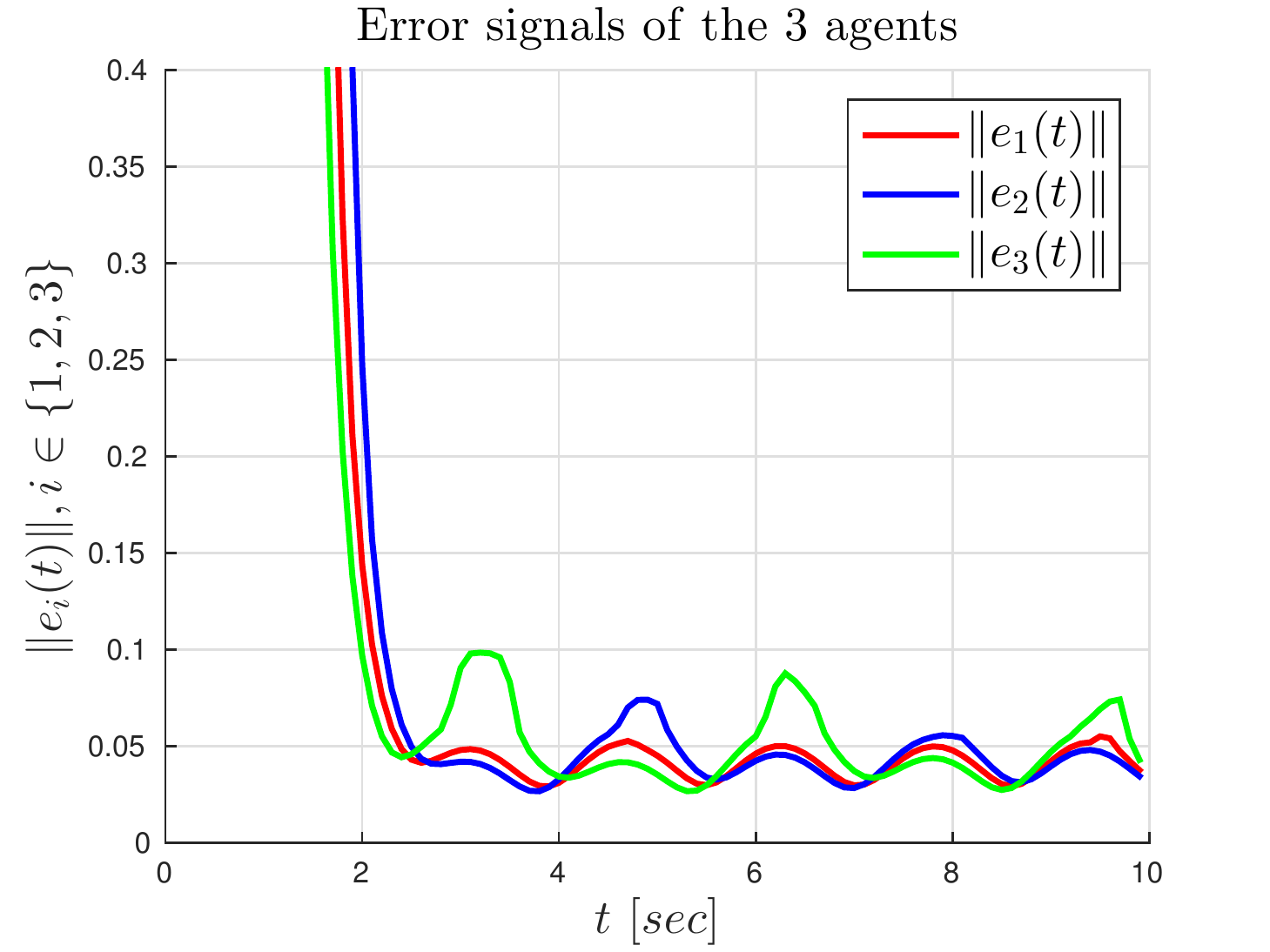}
	\caption{The evolution of the $2-$norms of the error signals of the three
			agents over time, in greater detail.}
	\label{fig:d_ON_res_3_2_errors_agent_1_zoom}
\end{figure}

\begin{figure}[t!]
	\centering
	\includegraphics[scale = 0.50]{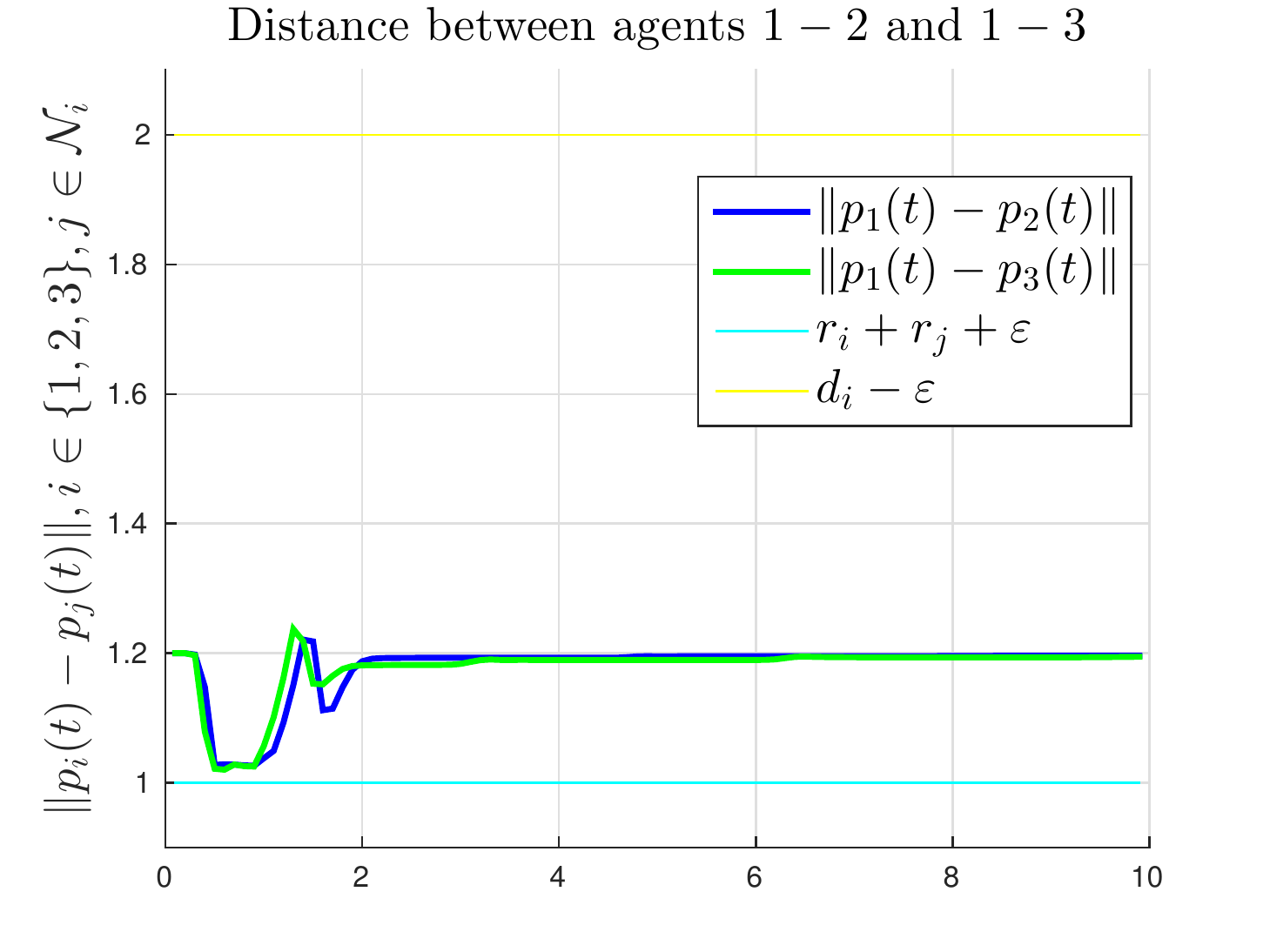}
	\caption{The distance between agents $1-2$ and $1-3$ over time. The maximum and the minimum allowed
			distances are $d_i-\varepsilon = 1.99$ and $r_i+r_j +\varepsilon = 1.01$, respectively for every $i \in \mathcal{V}$, $j \in \mathcal{N}_i$.}
	\label{fig:d_ON_res_3_2_distance_agents_13}
\end{figure}

\begin{figure}[t!]
	\centering
	\includegraphics[scale = 0.50]{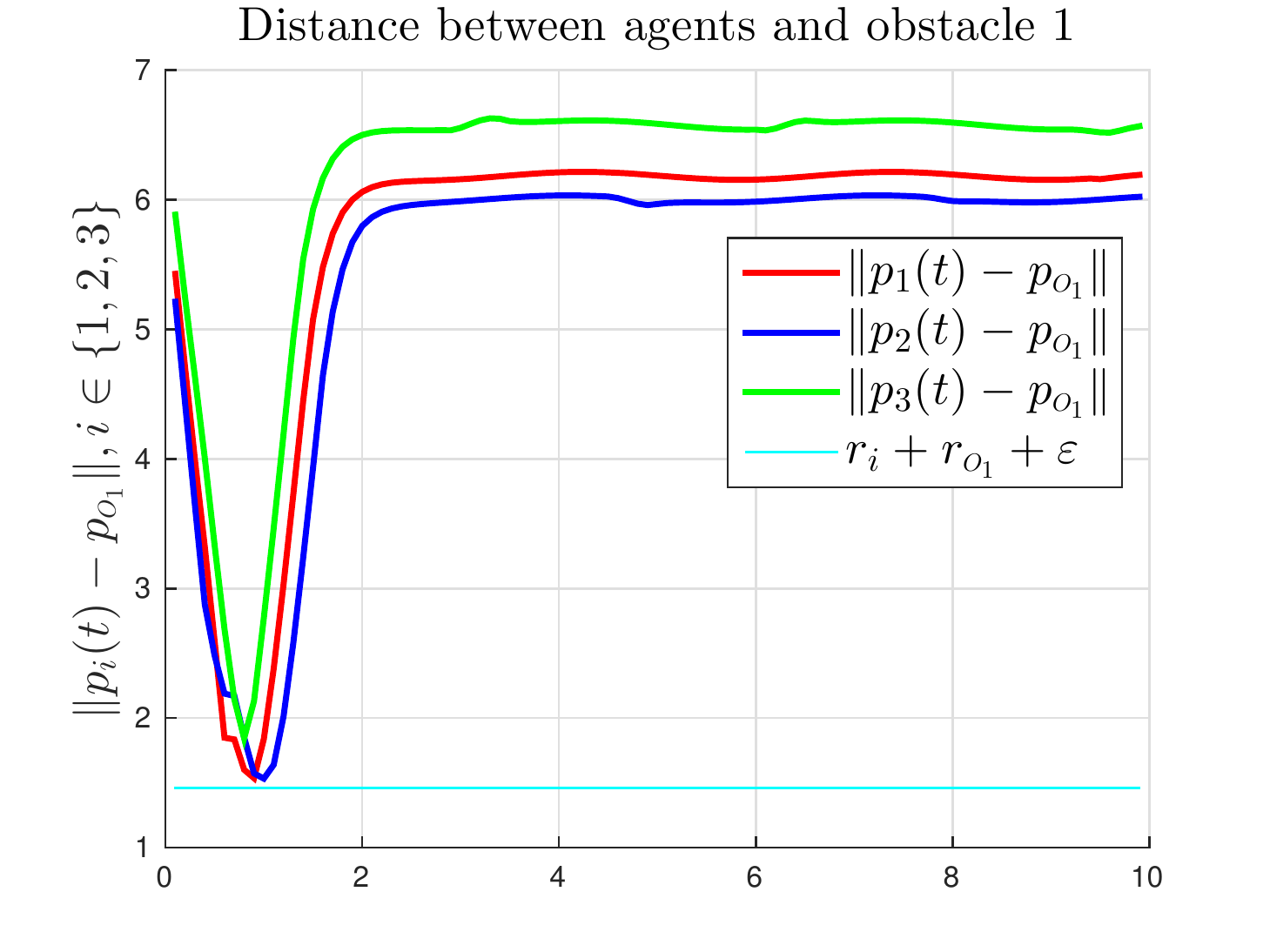}
	\caption{The distance between the agents and obstacle $1$ over time. The
			minimum allowed distance is $r_i + r_{\scriptscriptstyle O_1} + \varepsilon = 1.51$.}
	\label{fig:d_ON_res_3_2_distance_obstacle_1_agents}
\end{figure}

\begin{figure}[t!]
	\centering
	\includegraphics[scale = 0.50]{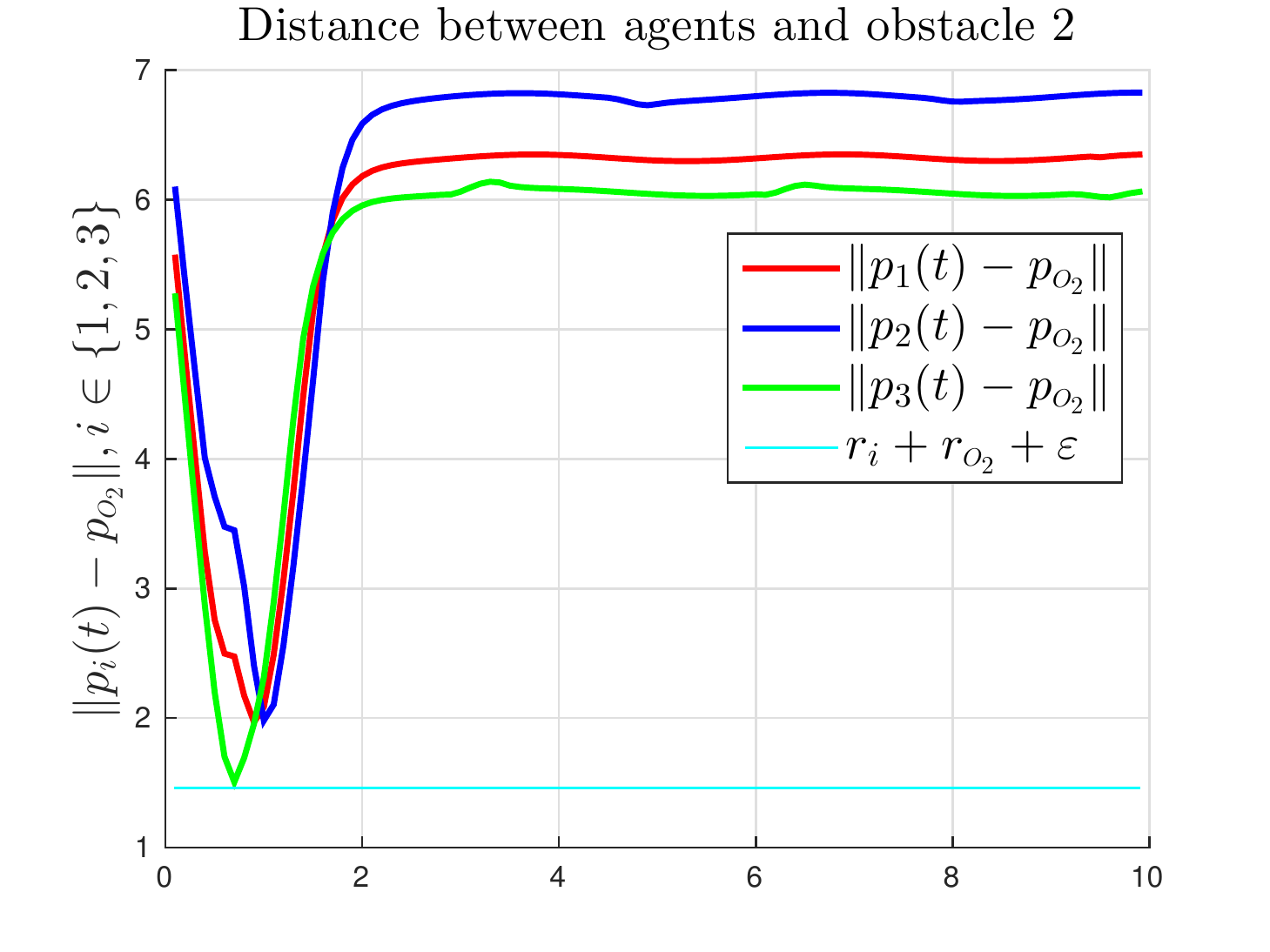}
	\caption{The distance between the agents and obstacle $2$ over time. The
			minimum allowed distance is $r_i + r_{\scriptscriptstyle O_2} + \varepsilon = 1.51$.}
	\label{fig:d_ON_res_3_2_distance_obstacle_2_agents}
\end{figure}

\begin{figure}[t!]
	\centering
	\includegraphics[scale = 0.50]{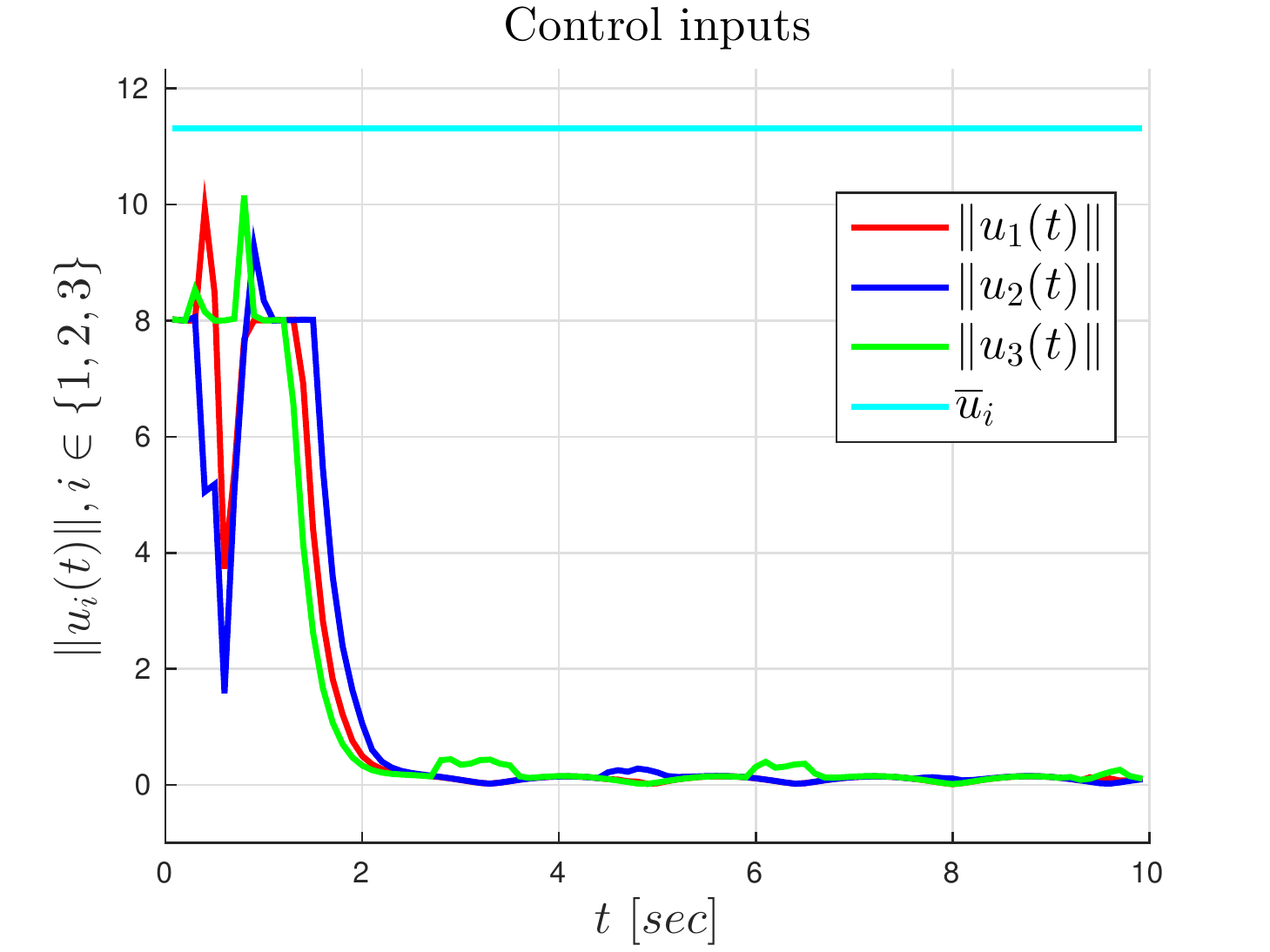}
	\caption{The norms of control input-signals directing the three agents over time. Their value is
			upper-bounded by $\overline{u}_i = 15$, as $\overline{u}_i$ defined in \eqref{eq:mathcal_WU}.}
	\label{fig:d_ON_res_3_2_inputs_agent_2}
\end{figure}

\begin{figure}[t!]
	\centering
	\includegraphics[scale = 0.50]{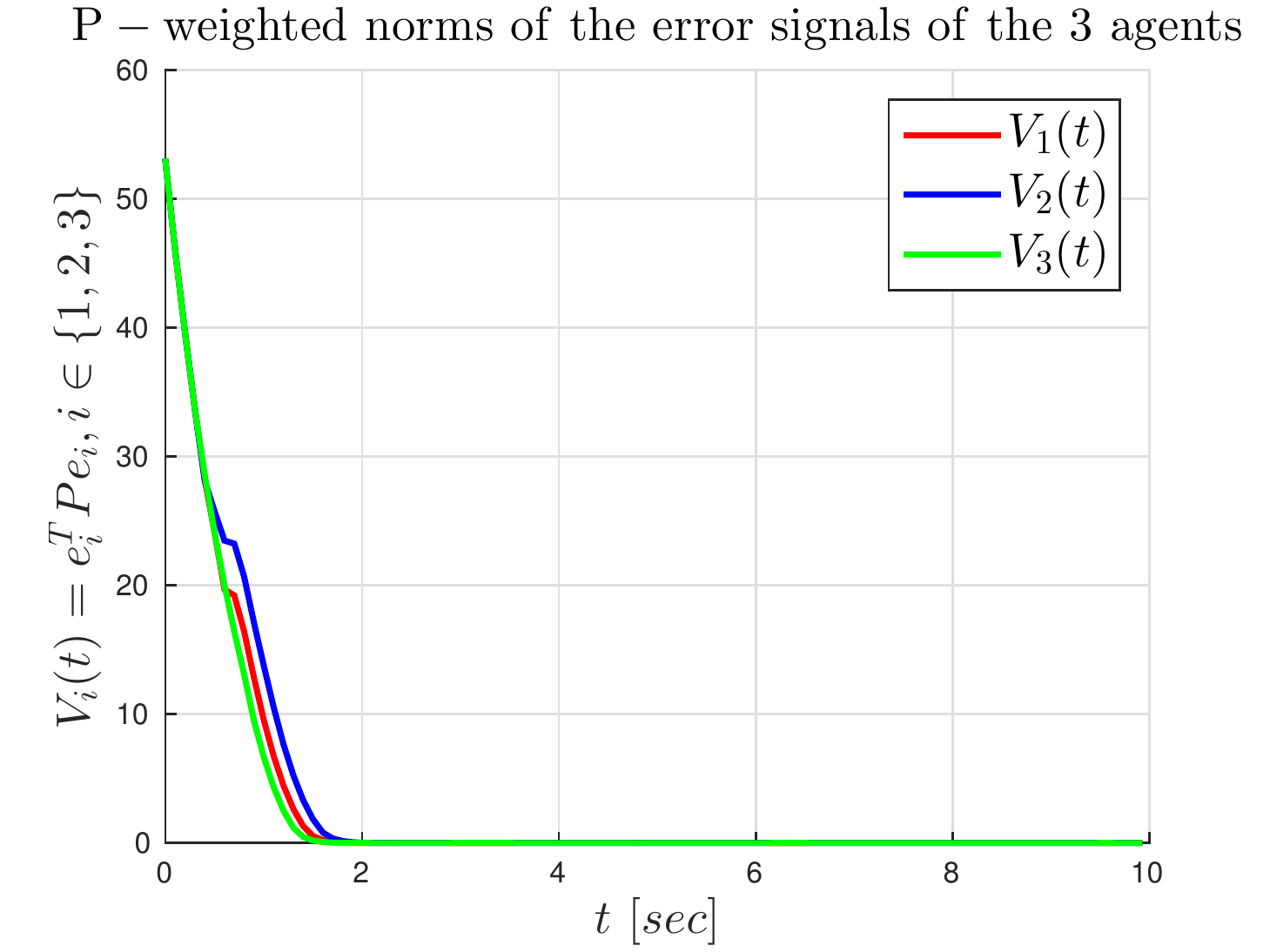}
	\caption{The $\mat{P}$-weighted norms of the errors of the three agents
			over time. $V_i(t), i = 1,2,3$ decreases monotonically until it reaches
			a value below the threshold $\varepsilon_{\Omega_i}$.}
	\label{fig:d_ON_res_3_2_V}
\end{figure}

\begin{figure}[t!]
	\centering
	\includegraphics[scale = 0.50]{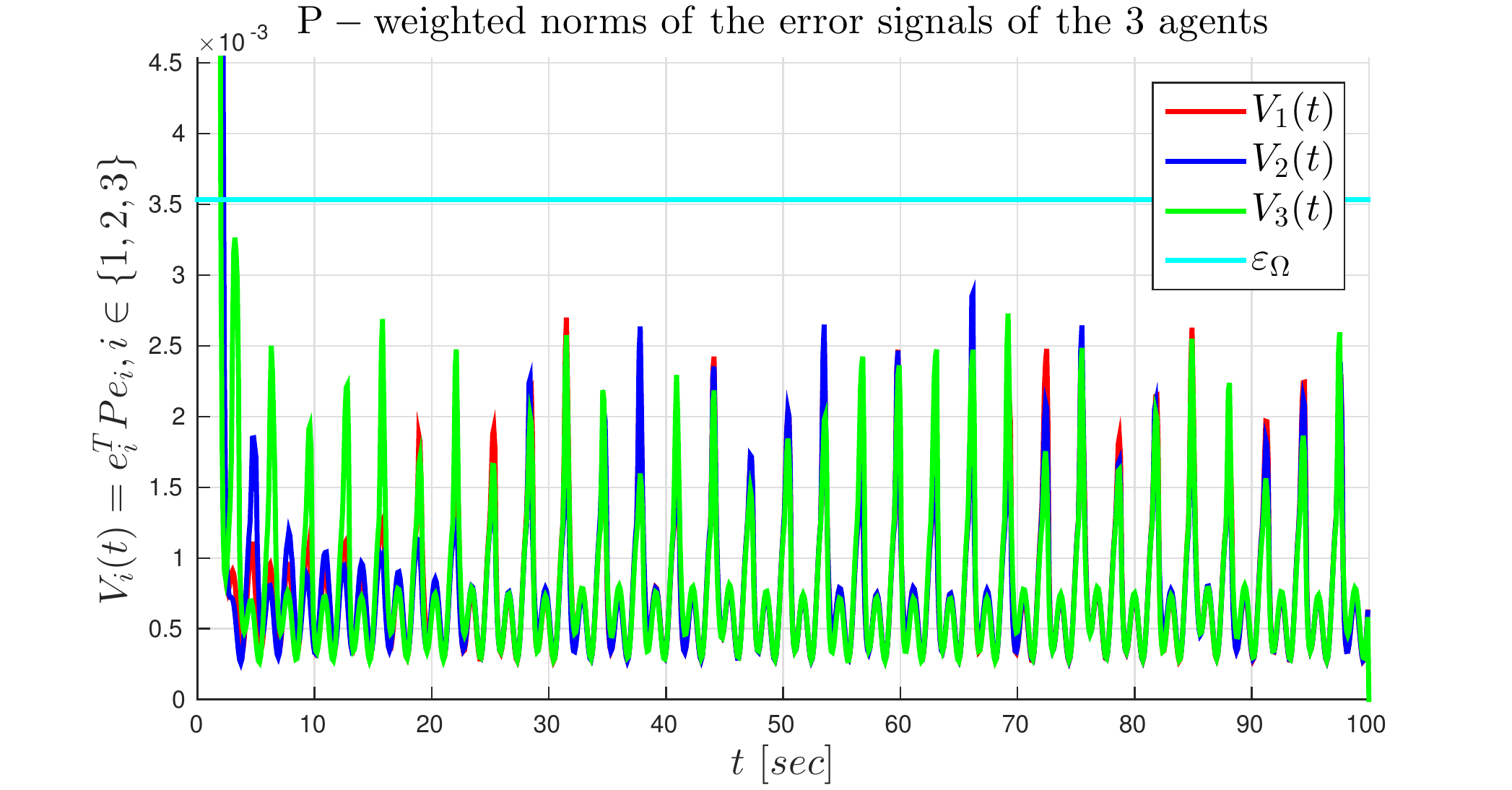}
	\caption{The $\mat{P}$-weighted norms of the errors of the three agents
			over an extended execution time of 100 seconds. Notice that the magnitudes
			of $V_i$ do not exceed the threshold $\varepsilon_{\Omega_i}, i = 1,2,3$
			as guaranteed by the control strategy. The periodic rise and fall of
			the magnitudes of $V_i$ is due to the periodic nature of the disturbance.}
	\label{fig:d_ON_res_3_2_V_zoom}
\end{figure}

The frames of the evolution of the trajectories of the three agents in the $x-y$
	plane are depicted in Figure \ref{fig:d_OFF_res_trajectory_3_2};
	Figure \ref{fig:d_ON_res_3_2_errors_agent_1} depicts the evolution of the
	error states' $2-$ norms of the agents; Figure \ref{fig:d_ON_res_3_2_errors_agent_1_zoom} depicts
	the evolution of the error states' $2-$ norms of the agents in greater detail;
	Figure \ref{fig:d_ON_res_3_2_distance_agents_13} shows the
	evolution of the distances between the neighboring agents;
	Figure \ref{fig:d_ON_res_3_2_distance_obstacle_1_agents} and
	Figure \ref{fig:d_ON_res_3_2_distance_obstacle_2_agents} depict the distance
	between the agents and the obstacle $1$ and $2$, respectively; Figure \ref{fig:d_ON_res_3_2_inputs_agent_2} shows the input signals directing
	the agents through time; Figure \ref{fig:d_ON_res_3_2_V} shows the
	evolution of the $\mat{P}$-norms of the errors of the three agents through
	time (i.e, $\vect{e}_i(t) \mat{P}_i \vect{e}_i(t)$, $i \in \{1,2,3\}$), and Figure \ref{fig:d_ON_res_3_2_V_zoom} shows the evolution of the
	$\mat{P}$-norms of the errors of the three agents through time in more detail,
	and for an extended execution time of $t=100$ seconds, without altering the rest
	of the simulation variables. Notably, the trajectories of the three agents
	are trapped inside the terminal set once they enter it, since the magnitudes of
	their $\mat{P}$-weighted error norms do not exceed the value of
	$\varepsilon_{\Omega_i}$ once they fall below it. Furthermore, it can be
	observed that all agents reach their desired goal by satisfying all the
	constraints imposed by Problem \ref{problem}. The simulation was performed in
MATLAB R2015a Environment utilizing the NMPC optimization routine provided in
\cite{grune2016nonlinear}. The simulation takes $1340 \sec$ on a desktop with
8 cores, 3.60GHz CPU and 16GB of RAM.

\section{Conclusions}
\label{sec:conclusions}
This paper addresses the problem of stabilizing a multiple rigid-bodies system
under constraints relating to the maintenance of connectivity between
agents, the aversion of collision among agents and between agents and
stationary obstacles within their working environment, and constraints
regarding their states and control inputs. The proposed framework is a Decentralized Nonlinear Model Predictive Control scheme. Simulation results verify the controller efficiency of the proposed framework. Future efforts will be devoted to reduce the communication
burden between the agents by introducing event-triggered communication controllers.

\section*{Disclosure Statement}
No potential conflict of interest was reported by the authors.

\section*{Funding}
This work was supported by the H2020 ERC Starting Grant BUCOPHSYS, the EU H2020 Co4Robots project, the Swedish Foundation for Strategic Research (SSF), the Swedish Research Council (VR) and the Knut och Alice Wallenberg Foundation (KAW).

\bibliographystyle{unsrt}
\bibliography{references}

\appendix

\section{Proof of Property \ref{property:set_property}}
\label{app:proof_set_property}
Consider the vectors $\vect{u}$, $\vect{v}$, $\vect{w}$, $\vect{x} \in \mathbb{R}^n$. According to Definition \ref{def:p_difference}, we have that: $\mathcal{S}_1 \ominus \mathcal{S}_2 = \{\vect{u} \in \mathbb{R}^n: \vect{u}+\vect{v} \in \mathcal{S}_1, \forall \ \vect{v} \in \mathcal{S}_2 \}$, $\mathcal{S}_2 \ominus \mathcal{S}_3 = \{\vect{w} \in \mathbb{R}^n: \vect{w}+\vect{x} \in \mathcal{S}_2, \forall \ \vect{x} \in \mathcal{S}_3 \}$.
Then, by adding the aforementioned sets according to Definition \ref{def:minkwoski} we get:
\begin{align}
(\mathcal{S}_1 \ominus \mathcal{S}_2) \oplus (\mathcal{S}_2 \ominus \mathcal{S}_3) & \notag \\
&\hspace{-18mm}= \{\vect{u} + \vect{w} \in \mathbb{R}^n : \vect{u} + \vect{v} \in \mathcal{S}_1 \ \text{and} \  \vect{w} + \vect{x} \in \mathcal{S}_2, \forall \ \vect{v} \in \mathcal{S}_2, \forall \ \vect{x} \in \mathcal{S}_3 \} \notag \\
&\hspace{-18mm}= \{\vect{u} + \vect{w} \in \mathbb{R}^n : \vect{u} + \vect{v} + \vect{w} + \vect{x} \in (\mathcal{S}_1 \oplus \mathcal{S}_2), \forall \ \vect{v} + \vect{x} \in (\mathcal{S}_2 \oplus \mathcal{S}_3) \}. \label{eq:u_v_w_x}
\end{align}
By setting $\vect{s}_1 = \vect{u} + \vect{w} \in \mathbb{R}^n$, $\vect{s}_2 = \vect{v} + \vect{x} \in \mathbb{R}^n$ and employing Definition \ref{def:p_difference}, \eqref{eq:u_v_w_x} becomes: $(\mathcal{S}_1 \ominus \mathcal{S}_2) \oplus (\mathcal{S}_2 \ominus \mathcal{S}_3)$ $= \{\vect{s}_1 \in \mathbb{R}^n : \vect{s}_1 + \vect{s}_2 \in (\mathcal{S}_1 \oplus \mathcal{S}_2)$, $\forall \ \vect{s}_2 \in (\mathcal{S}_2 \oplus \mathcal{S}_3) \}$ $= (\mathcal{S}_1 \oplus \mathcal{S}_2) \ominus (\mathcal{S}_2 \oplus \mathcal{S}_3)$,
which concludes the proof. \qed

\section{Proof of Property \ref{property 1}}
\label{app:proof_property_1}
By setting $\vect{z} = \vect{e} + \vect{z}_{\text{des}}$, $\vect{z}' = \vect{e}' + \vect{z}_{\text{des}}$ in \eqref{eq:lipschitz_f_i} we get: $\|f_i(\vect{e} + \vect{z}_{\text{des}}, \vect{u})-f_i(\vect{e}' + \vect{z}_{\text{des}}, \vect{u})\|$ $\le L_{f_i} \|\vect{e}  + \vect{z}_{\text{des}}-\vect{e}' - \vect{z}_{\text{des}}\|$. By using \eqref{eq:functions_g_i}, the latter becomes: $\|g_i(\vect{e}, \vect{u})-g_i(\vect{e}', \vect{u})\|$ $\le L_{g_i} \|\vect{e}-\vect{e}'\|$, where $L_{g_i} = L_{f_i}$, which leads to the conclusion of the proof. \qed

\section{Proof of Lemma \ref{lemma:F_i_bounded_K_class}}
\label{app:proof_lemma1}
By invoking the fact that:
\begin{align} \label{eq:rayleigh_inequality}
\lambda_{\min}(\mat{P}) \|\vect{y}\|^2 \le \vect{y}^\top \mat{P} \vect{y} \le \lambda_{\max}(\mat{P}) \|\vect{y}\|^2, \forall \vect{y} \in \mathbb{R}^n, \mat{P} \in \mathbb{R}^{n \times n}, \mat{P} = \mat{P}^\top > 0,
\end{align}
we have: $\vect{e}_i^\top \mat{Q}_i \vect{e}_i + \vect{u}_i^\top \mat{R}_i \vect{u}_i \le \max \{\lambda_{\max}(\mat{Q}_i), \lambda_{\max}(\mat{R}_i) \} \|\vect{\eta}_i \|^2$, and: $\vect{e}_i^\top \mat{Q}_i \vect{e}_i + \vect{u}_i^\top \mat{R}_i \vect{u}_i \ge \min \{\lambda_{\min}(\mat{Q}_i), \lambda_{\min}(\mat{R}_i) \} \|\vect{\eta}_i \|^2$, where $\vect{\eta}_i = \left[ \vect{e}_i^\top, \vect{u}_i^\top\right]^\top$ and $i \in \mathcal{V}$. By defining the $\mathcal{K}_{\infty}$ functions $\alpha_1$, $\alpha_2 : \mathbb{R}_{\ge 0}  \to \mathbb{R}_{\ge 0}$: $\alpha_1(y)$ $\triangleq \min \{\lambda_{\min}(\mat{Q}_i), \lambda_{\min}(\mat{R}_i) \} \|y \|^2$, $\alpha_2(y)$ $\triangleq \max \{\lambda_{\max}(\mat{Q}_i), \lambda_{\max}(\mat{R}_i) \} \|y \|^2$, we get $\alpha_1\big(\|\vect{\eta}_i\|\big) \leq F_i\big(\vect{e}_i, \vect{u}_i\big) \leq \alpha_2\big(\|\vect{\eta}_i \|\big)$. \hspace{-2mm} \qed

\section{Proof of Lemma \ref{lemma:F_Lipschitz}}
\label{app:proof_lemma2}
For every $\vect{e}_i, \vect{e}_i' \in \mathcal{E}_i$, and $\vect{u}_i \in \mathcal{U}_i$ it holds that:
\begin{align}
\big|F_i(\vect{e}_i, \vect{u}_i) - F_i(\vect{e}'_i, \vect{u}_i)\big|
&= \big|\vect{e}_i^{\top} \mat{Q}_i \vect{e}_i + \vect{u}_i^{\top} \mat{R}_i \vect{u}_i
-(\vect{e}_i')^{\top} \mat{Q}_i \vect{e}'_i - \vect{u}_i^{\top} \mat{R}_i \vect{u}_i \big| \notag \\
&\leq \big|\vect{e}_i^{\top} \mat{Q}_i (\vect{e}_i -\vect{e}'_i)\big| + \big| (\vect{e}'_i)^{\top} \mat{Q}_i (\vect{e}_i-\vect{e}'_i)\big|. \label{eq:inequality_lemma_2}
\end{align}
By employing the property that: $|\vect{e}_i^{\top} \mat{Q}_i \vect{e}_i'|$ $\leq \|\vect{e}_i \|  \| \mat{Q}_i \vect{e}_i'\|$ $\le  \| \mat{Q}_i \| \|\vect{e}_i \|  \| \vect{e}_i'\|$ $\le \sigma_{\max}(\mat{Q}_i) \|\vect{e}_i\| \|\vect{e}'_i\|$,
\eqref{eq:inequality_lemma_2} is written as: $\big|F_i(\vect{e}_i, \vect{u}_i) - F_i(\vect{e}_i', \vect{u}_i)\big|$ $\leq \sigma_{\max}(\mat{Q}_i) \|\vect{e}_i\| \|\vect{e}_i - \vect{e}_i'\|$ $+ \sigma_{\max}(\mat{Q}_i) \|\vect{e}_i'\| \|\vect{e}_i - \vect{e}_i'\|$ $\le \left[ 2 \sigma_{\max}(\mat{Q}_i) \sup \limits_{\vect{e}_i \in \mathcal{E}_i} \|\vect{e}_i\|  \right]  \|\vect{e}_i - \vect{e}_i'\|$ $= L_{F_i}  \|\vect{e}_i - \vect{e}_i'\|$. \qed

\section{Proof of Lemma \ref{lemma:diff_state_from_same_conditions}}
\label{proof:lemma_diff_state_from_same_conditions}
By employing Property \ref{remark:predicted_actual_equations_with_disturbance} and substituting $\tau \equiv t_k$ and $s \equiv t_k + s$ in \eqref{eq:remark_4_eq_1}, \eqref{eq:remark_4_eq_2} yields: $\vect{e}_i\big(t_k + s;\ \overline{\vect{u}}_i\big(\cdot;\ \vect{e}_i(t_k)\big), \vect{e}_i(t_k)\big)$ $= \vect{e}_i(t_k)$ $+ \int_{t_k}^{t_k + s} g_i\big(\vect{e}_i(s';\ \vect{e}_i(t_k)), \overline{\vect{u}}_i(s')\big) ds' + \int_{t_k}^{t_k + s}\vect{w}_i(\cdot, s')ds'$, $\overline{\vect{e}}_i\big(t_k + s;\ \overline{\vect{u}}_i\big(\cdot;\ \vect{e}_i(t_k)\big), \vect{e}_i(t_k)\big) =
\vect{e}_i(t_k)$ $+ \int_{t_k}^{t_k + s} g_i\big(\overline{\vect{e}}_i(s';\ \vect{e}_i(t_k)), \overline{\vect{u}}_i(s')\big) ds'$,
respectively. Subtracting the latter from the former and taking norms on both sides yields: $\bigg\|  \vect{e}_i\big(t_k + s;\ \overline{\vect{u}}_i\big(\cdot;\ \vect{e}_i(t_k)\big)$, $\vect{e}_i(t_k)\big)$ $- \overline{\vect{e}}_i\big(t_k + s;\ \overline{\vect{u}}_i\big(\cdot;\ \vect{e}_i(t_k)\big)$, $\vect{e}_i(t_k)\big) \bigg\|$ $=\bigg\| \int_{t_k}^{t_k + s} g_i\big(\vect{e}_i(s';\ \vect{e}_i(t_k)), \overline{\vect{u}}_i(s')\big) ds'$ $- \int_{t_k}^{t_k + s} g_i\big(\overline{\vect{e}}_i(s';\ \vect{e}_i(t_k))$, $\overline{\vect{u}}_i(s')\big) ds'$ $+ \int_{t_k}^{t_k + s} \vect{w}_i(\cdot, s')ds' \bigg\|$ $\leq L_{g_i} \int_{t_k}^{t_k + s} \bigg\| \vect{e}_i\big(s;\ \overline{\vect{u}}_i\big(\cdot;\ \vect{e}_i(t)\big)$, $\vect{e}_i(t)\big)$ $- \overline{\vect{e}}_i\big(s;\ \overline{\vect{u}}_i\big(\cdot;\ \vect{e}_i(t)\big)$, $\vect{e}_i(t)\big) \bigg\| ds$ $+ s \overline{w}_i$, since, according to Property \ref{property 1}, $g_i$ is locally Lipschitz continuous in $\mathcal{E}_i \times \mathcal{U}_i$ with Lipschitz constant $L_{g_i}$. Then, we get: $\bigg\| \vect{e}_i\big(t_k+s;\ \overline{\vect{u}}_i\big(\cdot;\ \vect{e}_i(t_k)\big)$, $\vect{e}_i(t_k)\big)$ $- \overline{\vect{e}}_i\big(t_k+s;\ \overline{\vect{u}}_i\big(\cdot;\ \vect{e}_i(t_k)\big)$, $\vect{e}_i(t_k)\big) \bigg\|$ $\leq s \overline{w}_i$ $+ L_{g_i} \int_{0}^{s} \bigg\| \vect{e}_i\big(t_k + s';\ \overline{\vect{u}}_i\big(\cdot;\ \vect{e}_i(t_k)\big)$, $\vect{e}_i(t_k)\big)$ $-
\overline{\vect{e}}_i\big(t_k + s';\ \overline{\vect{u}}_i\big(\cdot;\ \vect{e}_i(t_k)\big), \vect{e}_i(t_k)\big) \bigg\| ds'$. By applying the Gr\"{o}nwall-Bellman inequality (see \cite[Appendix A]{khalil_nonlinear_systems}) we get: $\bigg\| \vect{e}_i\big(t_k + s;\ \overline{\vect{u}}_i\big(\cdot;\ \vect{e}_i(t_k)\big), \vect{e}_i(t_k)\big) - \overline{\vect{e}}_i\big(t_k + s;\ \overline{\vect{u}}_i\big(\cdot;\ \vect{e}_i(t_k)\big), \vect{e}_i(t_k)\big) \bigg\|$ $\le \dfrac{\overline{w}_i}{L_{g_i}} (e^{L_{g_i}s} - 1)$. \hspace{-3mm} \qed

\section{Proof of Property \ref{property:restricted_constraint_set}}
\label{app:property_2}
Let us define the function $\vect{\zeta}_i : \mathbb{R}_{\geq 0} \to \mathcal{M} \times \mathbb{R}^{6}$ as:
$\vect{\zeta}_i(s) \triangleq \vect{e}_i(s) - \overline{\vect{e}}_i(s;\ \vect{u}_i(s;\ \vect{e}_i(t_k)), \vect{e}_i(t_k))$,
for $s \in [t_k, t_k + T_p]$. According to Lemma \ref{lemma:diff_state_from_same_conditions} we have that: $\|\vect{\zeta}_i(s)\|$ $= \|\vect{e}_i(s) - \overline{\vect{e}}_i\big(s;\ \vect{u}_i(s;\ \vect{e}_i(t)), \vect{e}_i(t)\big)\|$ $\leq \dfrac{\overline{w}_i}{L_{g_i}} (e^{L_{g_i} (s-t)} - 1), s \in [t_k, t_k + T_p]$, which means that $\vect{\zeta}_i(s) \in \mathcal{X}_{i,s-t}$. Now we have that: $\overline{\vect{e}}_i\big( s;\ \vect{u}_i(\cdot,\ \vect{e}_i(t_k)), \vect{e}_i(t_k) \big) \in \mathcal{E}_i \ominus \mathcal{X}_{i,s-t_k}$.
Then, it holds that: $\vect{\zeta}_i(s)$ $+ \overline{\vect{e}}_i\big(s;\ \vect{u}_i(s;\ \vect{e}_i(t_k)), \vect{e}_i(t_k)\big)$ $\in \big(\mathcal{E}_i \ominus \mathcal{X}_{i,s-t_k}\big) \oplus \mathcal{X}_{i,s-t_k}$, or: $\vect{e}_i(s) \in \big(\mathcal{E}_i \ominus \mathcal{X}_{i,s-t_k}\big)$ $\oplus \mathcal{X}_{i,s-t_k}$. Theorem 2.1 (ii) from \cite{kolmanovsky} states that for every $U,V \subseteq \mathbb{R}^n$ it holds that: $\left(U \ominus V \right) \oplus V \subseteq U$. By invoking the latter result we get: $\vect{e}_i(s) \in \big(\mathcal{E}_i$ $\ominus \mathcal{X}_{i,s-t_k}\big) \oplus \mathcal{X}_{i,s-t_k} \subseteq \mathcal{E}_i$ $\Rightarrow \vect{e}_i(s) \in \mathcal{E}_i$, $s \in [t_k, t_k+T_p]$. \qed

\section{Proof of Lemma \ref{lemma:V_i_lower_upper_bounded}}
\label{appendix_lemma_4}
By invoking \eqref{eq:rayleigh_inequality} we get: $\lambda_{\min}(\mat{P}_i) \|\vect{e}_i\|^2 \le \vect{e}_i^\top \mat{P}_i \vect{e}_i \le \lambda_{\max}(\mat{P}_i) \|\vect{e}_i\|^2$, $\forall \vect{e}_i \in \Psi_i$, $i \in \mathcal{V}$. By defining the $\mathcal{K}_{\infty}$ functions $\alpha_1$, $\alpha_2 : \mathbb{R}_{\ge 0}  \to \mathbb{R}_{\ge 0}$: $\alpha_1(y) \triangleq \lambda_{\min}(\mat{P}_i) \|y \|^2, \alpha_2(y) \triangleq \lambda_{\max}(\mat{P}_i) \|y \|^2$, we get: $\alpha_1\big(\|\vect{e}_i\|\big) \leq V_i(\vect{e}_i) \leq \alpha_2\big(\| \vect{e}_i \|\big)$, $\forall \vect{e}_i \in \Psi_i$, $i \in \mathcal{V}$. \qed

\section{Feasibility Analysis}
\label{app:feasibility}
In this section we will show that there can be constructed an admissible but not necessarily optimal control input according to Definition \ref{definition:admissible_input_with_disturbance}.

Consider a sampling instant $t_k$ for which a solution $\overline{\vect{u}}_i^{\star}\big(\cdot;\ \vect{e}_i(t_k)\big)$ to Problem $1$ exists. Suppose now a time instant $t_{k+1}$ such that $t_k < t_{k+1} < t_k + T_p$, and consider that the
optimal control signal calculated at $t_k$ is comprised by the following two
portions:
\begin{equation}
\overline{\vect{u}}_i^{\star}\big(\cdot;\ \vect{e}_i(t_k)\big) = \left\{
\begin{array}{ll}
\overline{\vect{u}}_i^{\star}\big(\tau_1;\ \vect{e}_i(t_k)\big), & \tau_1 \in [t_k, t_{k+1}], \\
\overline{\vect{u}}_i^{\star}\big(\tau_2;\ \vect{e}_i(t_k)\big), & \tau_2 \in [t_{k+1}, t_k + T_p].
\end{array}
\right.
\label{eq:optimal_input_portions_with_disturbances}
\end{equation}

Both portions are admissible since the calculated optimal control input is
admissible, and hence they both conform to the input constraints.
As for the resulting predicted states, they satisfy the state constraints, and,
crucially:
\begin{align}
\overline{\vect{e}}_i\big(t_k + T_p;\ \overline{\vect{u}}_i^{\star}(\cdot), \vect{e}_i(t_k)\big) \in \Omega_i.
\label{eq:predicted_t_k_T_p_from_t_k_in_omega}
\end{align}
Furthermore, according to condition $(3)$ of Theorem \ref{theorem:with_disturbances}, there exists an admissible (and certainly not guaranteed optimal feedback control) input $\kappa_i \in \mathcal{U}_i$ that renders $\Psi_i$ (and consequently $\Omega_i$) invariant over $[t_k + T_p, t_{k+1} + T_p]$.

Given the above facts, we can construct an admissible input $\widetilde{\vect{u}}_i(\cdot)$  for time $t_{k+1}$ by sewing together the second portion of \eqref{eq:optimal_input_portions_with_disturbances} and the admissible input $\kappa_i(\cdot)$:

\begin{equation}
\widetilde{\vect{u}}_i(\tau) = \left\{
\begin{array}{ll}
\overline{\vect{u}}_i^{\star}\big(\tau;\ \vect{e}_i(t_k)\big), & \tau \in [t_{k+1}, t_k + T_p], \\
\kappa_i \big(\overline{\vect{e}}_i\big(\tau;\ \overline{\vect{u}}_i^{\star}(\cdot), \vect{e}_i(t_{k+1})\big)\big), & \tau \in (t_k + T_p, t_{k+1} + T_p].
\end{array}
\right.
\label{eq:optimal_input_t_plus_one_with_disturbances}
\end{equation}

Applied at time $t_{k+1}$, $\widetilde{\vect{u}}_i(\tau)$
is an admissible control input with respect to the input constraints as a composition of admissible control inputs, for all $\tau \in [t_{k+1}, t_{k+1} + T_p]$. What remains to prove is the following two statements:\\

\noindent \textbf{Statement 1 :} $\vect{e}_i\big(t_{k+1} + s;\ \overline{\vect{u}}_i^{\star}(\cdot), \vect{e}_i(t_{k+1})\big) \in \mathcal{E}_i$, $\forall s \in [0,T_p]$.

\noindent \textbf{Statement 2 :} $\overline{\vect{e}}_i\big(t_{k+1} + T_p;\ \widetilde{\vect{u}}_i(\cdot), \vect{e}_i(t_{k+1}) \big) \in \Omega_i$. \\

\noindent \textbf{Proof of Statement 1 :} Initially we have that: $\overline{\vect{e}}_i\big(t_{k+1} + s;\ \widetilde{\vect{u}}_i(\cdot), \vect{e}_i(t_{k+1})\big) \in \mathcal{E}_i \ominus \mathcal{X}_{s}$,
for all $s \in [0, T_p]$. By applying Lemma \ref{lemma:diff_state_from_same_conditions} for $t=t_{k+1} + s$ and $\tau=t_k$ we get $\bigg\| \vect{e}_i\big(t_{k+1}+s;\  \overline{\vect{u}}_i^{\star}(\cdot), \vect{e}_i(t_k)\big)-\overline{\vect{e}}_i\big(t_{k+1} + s;\ \overline{\vect{u}}_i^{\star}(\cdot), \vect{e}_i(t_k)\big) \bigg\|$ $\leq \dfrac{\overline{w}_i}{L_{g_i}}\big(e^{L_{g_i} (h+s)}-1\big)$, or equivalently: $\vect{e}_i\big(t_{k+1}+s;\  \overline{\vect{u}}_i^{\star}(\cdot), \vect{e}_i(t_k)\big)$ $-\overline{\vect{e}}_i\big(t_{k+1} + s;\ \overline{\vect{u}}_i^{\star}(\cdot), \vect{e}_i(t_k)\big) \in \mathcal{X}_{i, h+s}$. By applying a reasoning identical to the proof of Lemma \ref{lemma:diff_state_from_same_conditions} for $t=t_{k+1}$ (in the model equation) and $t = t_k$ (in the real model equation), and $\tau = s$ we get: $\bigg\|\vect{e}_i\big(t_{k+1}+s;\  \overline{\vect{u}}_i^{\star}(\cdot), \vect{e}_i(t_k)\big)$ $-\overline{\vect{e}}_i\big(t_{k+1} + s;\ \overline{\vect{u}}_i^{\star}(\cdot), \vect{e}_i(t_{k+1})\big)\bigg\|$ $\leq \dfrac{\overline{w}_i}{L_{g_i}}\big(e^{L_{g_i} s}-1\big)$, which translates to: $\vect{e}_i\big(t_{k+1}+s;\  \overline{\vect{u}}_i^{\star}(\cdot), \vect{e}_i(t_k)\big)$ $-\overline{\vect{e}}_i\big(t_{k+1} + s;\ \overline{\vect{u}}_i^{\star}(\cdot), \vect{e}_i(t_{k+1})\big) \in \mathcal{X}_{i,s}$.

Furthermore, we know that the solution to the optimization problem is
feasible at time $t_k$, which means that: $\overline{\vect{e}}_i\big(t_{k+1} + s;\ \overline{\vect{u}}_i^{\star}(\cdot), \vect{e}_i(t_k)\big) \in \mathcal{E}_i \ominus \mathcal{X}_{i,h+s}$. Let us for sake of readability set: $\vect{e}_{i, 0}$ $= \vect{e}_i\big(t_{k+1} + s;\ \overline{\vect{u}}_i^{\star}(\cdot), \vect{e}_i(t_k)\big)$, $\overline{\vect{e}}_{i, 0}$ $= \overline{\vect{e}}_i\big(t_{k+1} + s;\ \overline{\vect{u}}_i^{\star}(\cdot), \vect{e}_i(t_k)\big)$, $\overline{\vect{e}}_{i, 1}$ $= \overline{\vect{e}}_i\big(t_{k+1} + s;\ \overline{\vect{u}}_i^{\star}(\cdot), \vect{e}_i(t_{k+1})\big)$, and translate the above system of inclusion relations: $\vect{e}_{i,0} - \overline{\vect{e}}_{i,0} \in \mathcal{X}_{i, h+s}$, $\vect{e}_{i,0} - \overline{\vect{e}}_{i,1} \in \mathcal{X}_{i,s}$, $\overline{\vect{e}}_{i,0} \in \mathcal{E}_i \ominus \mathcal{X}_{i,h+s}$.

First we will focus on the first two relations, and we will derive a result that will combine with the third statement so as to prove that the predicted
state will be feasible from $t_{k+1}$ to $t_{k+1} + T_p$. Subtracting the second from the first yields $\overline{\vect{e}}_{i,1} - \overline{\vect{e}}_{i,0} \in \mathcal{X}_{i, h+s} \ominus \mathcal{X}_{i,s}$. Now we use the third relation $\overline{\vect{e}}_{i,0} \in \mathcal{E}_i \ominus \mathcal{X}_{i,h+s}$, along with: $\overline{\vect{e}}_{i,1}- \overline{\vect{e}}_{i,0} \in \mathcal{X}_{i, h+s} \ominus \mathcal{X}_{i,s}$. Adding the latter to the former yields: $\overline{\vect{e}}_{i,1} \in \big(\mathcal{E}_i \ominus \mathcal{X}_{i,h+s}\big) \oplus \big(\mathcal{X}_{i, h+s} \ominus \mathcal{X}_{i,s}\big)$. By using Property \ref{property:set_property} we get: $\overline{\vect{e}}_{i,1} \in \big(\mathcal{E}_i \oplus \mathcal{X}_{i,h+s}\big) \ominus \big(\mathcal{X}_{i, h+s} \oplus \mathcal{X}_{i,s}\big)$. Using implication \footnote{$A = B_1 \oplus B_2 \Rightarrow A \ominus B = (A \ominus B_1) \ominus B_2$}
(v) of Theorem 2.1 from \cite{kolmanovsky} yields: $\overline{\vect{e}}_{i,1} \in \bigg(\big(\mathcal{E}_i \oplus \mathcal{X}_{i,h+s}\big) \ominus \mathcal{X}_{i, h+s}\bigg) \ominus \mathcal{X}_{i,s}$. Using implication \footnote{$(A \oplus B) \ominus B \subseteq A$} (3.1.11) from \cite{schneider_2013} yields $\overline{\vect{e}}_{i,1} \in \mathcal{E}_i \ominus \mathcal{X}_{i,s}$, or equivalently:
\begin{align}
\overline{\vect{e}}_i\big(t_{k+1} + s;\ \overline{\vect{u}}_i^{\star}(\cdot), \vect{e}_i(t_{k+1})\big) \in \mathcal{E}_i \ominus \mathcal{X}_{i,s},\
\forall s \in [0,T_p].
\label{eq:feasibility_2}
\end{align}

By consulting with Property \ref{property:restricted_constraint_set}, this means that the state of the ``true" system does not violate the constraints $\mathcal{E}_i$ over the horizon $[t_{k+1}, t_{k+1} + T_p]$: $\overline{\vect{e}}_i\big(t_{k+1}$ $+ s;\ \overline{\vect{u}}_i^{\star}(\cdot)$, $\vect{e}_i(t_{k+1})\big)$ $\in \mathcal{E}_i$ $\ominus \mathcal{X}_{i,s}$ $\Rightarrow \vect{e}_i\big(t_{k+1} + s;\ \overline{\vect{u}}_i^{\star}(\cdot)$, $\vect{e}_i(t_{k+1})\big) \in \mathcal{E}_i$, $\forall s \in [0,T_p]$.

\noindent \textbf{Proof of Statement 3}: To prove this statement we begin with:
\begin{align}
V_i\big(\overline{\vect{e}}_i\big(t_k + T_p;&\ \overline{\vect{u}}_i^{\star}(\cdot), \vect{e}_i(t_{k+1})\big)\big)
- V_i\big(\overline{\vect{e}}_i\big(t_k + T_p;\ \overline{\vect{u}}_i^{\star}(\cdot), \vect{e}_i(t_k)\big) \notag \\
&\hspace{2mm}\leq L_{V_i}\bigg \| \overline{\vect{e}}_i\big(t_k + T_p;\ \overline{\vect{u}}_i^{\star}(\cdot), \vect{e}_i(t_{k+1})\big)
- \overline{\vect{e}}_i\big(t_k + T_p;\ \overline{\vect{u}}_i^{\star}(\cdot), \vect{e}_i(t_k) \big)\Big\|. \label{eq:from_DV_to_De}
\end{align}
Consulting with Remark \ref{remark:predicted_actual_equations_with_disturbance} we get that the two terms inside the norm are respectively equal to: $\overline{\vect{e}}_i\big(t_k + T_p;\ \overline{\vect{u}}_i^{\star}(\cdot), \vect{e}_i(t_{k+1})\big) = \vect{e}_i(t_{k+1}) + \int_{t_{k+1}}^{t_k + T_p} g_i\big(\overline{\vect{e}}_i(s;\ \vect{e}_i(t_{k+1})), \overline{\vect{u}}_i^{\star}(s) \big)ds$, and $\overline{\vect{e}}_i\big(t_k + T_p;\ \overline{\vect{u}}_i^{\star}(\cdot), \vect{e}_i(t_k)\big)$ $= \overline{\vect{e}}_i(t_{k+1})$ $+ \int_{t_{k+1}}^{t_k + T_p} g_i\big(\overline{\vect{e}}_i(s;\ \vect{e}_i(t_k)), \overline{\vect{u}}_i^{\star}(s) \big)ds$. Subtracting the latter from the former and taking norms on both sides we get: $\bigg \|\overline{\vect{e}}_i\big(t_k + T_p;\ \overline{\vect{u}}_i^{\star}(\cdot)$, $\vect{e}_i(t_{k+1})\big)$ $- \overline{\vect{e}}_i\big(t_k + T_p;\ \overline{\vect{u}}_i^{\star}(\cdot)$, $\vect{e}_i(t_k)\big) \bigg\|$ $\le \bigg \| \vect{e}_i(t_{k+1})$ $- \overline{\vect{e}}_i(t_{k+1}) \bigg\|$ $+  L_{g_i} \int_{h}^{T_p} \bigg\| \overline{\vect{e}}_i\big(t_k + s;\ \overline{\vect{u}}_i^{\star}(\cdot)$, $\vect{e}_i(t_{k+1})\big)$ $- \overline{\vect{e}}_i\big(t_k + s;\ \overline{\vect{u}}_i^{\star}(\cdot)$, $\vect{e}_i(t_k)\big) \bigg\| ds$. By applying the  Gr\"{o}nwall-Bellman inequality we obtain: $\bigg \|\overline{\vect{e}}_i\big(t_k + T_p;\ \overline{\vect{u}}_i^{\star}(\cdot)$, $\vect{e}_i(t_{k+1})\big)$ $- \overline{\vect{e}}_i\big(t_k + T_p; \ \overline{\vect{u}}_i^{\star}(\cdot)$, $\vect{e}_i(t_k)\big) \bigg\|$ $\leq \bigg \| \vect{e}_i(t_{k+1})$ $- \overline{\vect{e}}_i(t_{k+1}) \bigg\| e^{L_{g_i} (T_p - h)}$. By applying Lemma \ref{lemma:diff_state_from_same_conditions} for $t = t_k$ and $\tau = h$ we have: $\bigg \|\overline{\vect{e}}_i\big(t_k + T_p;\ \overline{\vect{u}}_i^{\star}(\cdot)$, $\vect{e}_i(t_{k+1})\big)$ $- \overline{\vect{e}}_i\big(t_k + T_p;\ \overline{\vect{u}}_i^{\star}(\cdot)$, $\vect{e}_i(t_k)\big) \bigg\|$ $\leq \dfrac{\overline{w}_i}{L_{g_i}} (e^{L_{g_i}h} - 1) e^{L_{g_i} (T_p - h)}$.

\noindent Hence \eqref{eq:from_DV_to_De} becomes:
\begin{align}
V_i\big(\overline{\vect{e}}_i\big(&t_k + T_p;\ \overline{\vect{u}}_i^{\star}(\cdot), \vect{e}_i(t_{k+1})\big)\big)
- V_i\big(\overline{\vect{e}}_i\big(t_k + T_p;\ \overline{\vect{u}}_i^{\star}(\cdot), \vect{e}_i(t_k)\big) \notag \\
&\hspace{68mm}= L_{V_i} \dfrac{\overline{w}_i}{L_{g_i}} (e^{L_{g_i}h} - 1) e^{L_{g_i} (T_p - h)}.
\label{eq:from_DV_to_eq}
\end{align}

Since the solution to the optimization problem is assumed to be feasible at time $t_k$, all states fulfill their respective constraints, and in particular, from \eqref{eq:predicted_t_k_T_p_from_t_k_in_omega}, the predicted state $\overline{\vect{e}}_i\big(t_k + T_p;\ \overline{\vect{u}}_i^{\star}(\cdot), \vect{e}_i(t_k)\big) \in \Omega_i$. This means that $V_i\big(\overline{\vect{e}}_i\big(t_k + T_p;\ \overline{\vect{u}}_i^{\star}(\cdot), \vect{e}_i(t_k)\big) \leq \varepsilon_{\Omega_i}$. Hence \eqref{eq:from_DV_to_eq} becomes: $V_i\big(\overline{\vect{e}}_i\big(t_k + T_p;\ \overline{\vect{u}}_i^{\star}(\cdot), \vect{e}_i(t_{k+1})\big)\big)$ $\leq \varepsilon_{\Omega_i}$ $+ L_{V_i} \dfrac{\overline{w}_i}{L_{g_i}} (e^{L_{g_i}h} - 1) e^{L_{g_i} (T_p - h)}$. From Assumption 4 of Theorem \ref{theorem:with_disturbances}, the upper bound of the disturbance is in turn bounded by: $\overline{w}_i \leq \dfrac{\varepsilon_{\Psi_i} - \varepsilon_{\Omega_i}}{\dfrac{L_{V_i}}{L_{g_i}} (e^{L_{g_i}h} - 1) e^{L_{g_i} (T_p - h)}}$. Therefore: $V_i\big(\overline{\vect{e}}_i\big(t_k + T_p;\ \overline{\vect{u}}_i^{\star}(\cdot), \vect{e}_i(t_{k+1})\big)\big) \leq \varepsilon_{\Psi_i}$, or, expressing the above in terms of $t_{k+1}$ instead of $t_k$: $V_i\big(\overline{\vect{e}}_i\big(t_{k+1} + T_p-h;\ \overline{\vect{u}}_i^{\star}(\cdot), \vect{e}_i(t_{k+1})\big)\big)$ $\leq \varepsilon_{\Psi_i}$. This means that the state
$\overline{\vect{e}}_i\big(t_{k+1} + T_p-h;\ \overline{\vect{u}}_i^{\star}(\cdot), \vect{e}_i(t_{k+1})\big) \in \Psi_i$.
From Assumption \ref{ass:psi_omega}, and since $\Psi_i \subseteq \Phi_i$, there is an admissible control signal
$\kappa_i \big(\overline{\vect{e}}_i\big(t_{k+1} + T_p-h;\ \overline{\vect{u}}_i^{\star}(\cdot), \vect{e}_i(t_{k+1})\big)\big)$
such that: $\overline{\vect{e}}_i\big(t_{k+1} + T_p;\ \kappa_i(\cdot), \overline{\vect{e}}_i\big(t_{k+1} + T_p - h;\ \overline{\vect{u}}_i^{\star}(\cdot), \vect{e}_i(t_{k+1})\big)\big) \in \Omega_i$. Hence, overall, it holds that:
\begin{align}
\overline{\vect{e}}_i\big(t_{k+1} + T_p;\ \widetilde{\vect{u}}_i(\cdot), \vect{e}_i(t_{k+1}) \big) \in \Omega_i.
\label{eq:feasibility_3}
\end{align}

Piecing the admissibility of $\widetilde{\vect{u}}_i(\cdot)$ from \eqref{eq:optimal_input_t_plus_one_with_disturbances} together with conclusions \eqref{eq:feasibility_2} and \eqref{eq:feasibility_3}, we conclude that the
application of the control input $\widetilde{\vect{u}}_i(\cdot)$ at time
$t_{k+1}$ results in that the states of the real system fulfill their intended constraints during the entire horizon $[t_{k+1}, t_{k+1} + T_p]$. Therefore,
overall, the (sub-optimal) control input $\widetilde{\vect{u}}_i(\cdot)$ is
admissible at time $t_{k+1}$ according to Definition
\ref{definition:admissible_input_with_disturbance}, which means
that feasibility of a solution to the optimization problem at time $t_k$ implies
feasibility at time $t_{k+1} > t_k$. Thus, since at time $t=0$ a solution is
assumed to be feasible, a solution to the optimal control problem is feasible
for all $t \geq 0$. \qed

\section{Convergence Analysis}
\label{app:convergence}
The second part of the proof involves demonstrating that the state $\vect{e}_i$ is ultimately bounded in $\Omega_i$. We will show that the
\textit{optimal} cost $J_i^{\star}\big(\vect{e}_i(t)\big)$ is an ISS Lyapunov function for the closed loop system \eqref{eq:error_system_perturbed}, under the control input \eqref{eq:position_based_optimal_u_2}, where: $J_i^{\star}\big(\vect{e}_i(t)\big) \triangleq J_i \Big(\vect{e}_i(t), \overline{\vect{u}}_i^{\star}\big(\cdot;\ \vect{e}_i(t)\big)\Big).$ For notational convenience, let us as define the following terms:

\begin{itemize}
	\item $\vect{u}_{0,i}(\tau) \triangleq \overline{\vect{u}}_i^{\star}\big(\tau;\ \vect{e}_i(t_k)\big)$
	as the \textit{optimal} input that results from the solution to Problem $1$ based on the measurement of state
	$\vect{e}_i(t_k)$, applied at time $\tau \geq t_k$;
	\item $\vect{e}_{0,i}(\tau) \triangleq \overline{\vect{e}}_i\big(\tau;\ \overline{\vect{u}}_i^{\star}\big(\cdot;\ \vect{e}_i(t_k)\big), \vect{e}_i(t_k)\big)$
	as the \textit{predicted} state at time $\tau \geq t_k$, that is, the predicted state that results from the application of the above input
	$\overline{\vect{u}}_i^{\star}\big(\cdot;\ \vect{e}_i(t_k)\big)$ to the
	state $\vect{e}_i(t_k)$, at time $\tau$;
	\item $\vect{u}_{1,i}(\tau) \triangleq \widetilde{\vect{u}}_i(\tau)$
	as the \textit{admissible} input at $\tau \geq t_{k+1}$
	(see \eqref{eq:optimal_input_t_plus_one_with_disturbances});
	\item $\vect{e}_{1,i}(\tau) \triangleq \overline{\vect{e}}_i\big(\tau;\ \widetilde{\vect{u}}_i(\cdot), \vect{e}_i(t_{k+1})\big)$
	as the \textit{predicted} state at time $\tau \geq t_{k+1}$, that is,
	the predicted state that results from the application of the above input
	$\widetilde{\vect{u}}_i(\cdot)$ to the state
	$\vect{e}_i\big(t_{k+1};\ \overline{\vect{u}}_i^{\star}\big(\cdot;\ \vect{e}_i(t_k)\big), \vect{e}_i(t_k)\big)$, at time $\tau$.
\end{itemize}

Before beginning to prove convergence, it is worth noting that while the cost $J_i \Big(\vect{e}_i(t), \overline{\vect{u}}_i^{\star}\big(\cdot;\ \vect{e}_i(t)\big)\Big),$ is optimal (in the sense that it is based on the optimal input, which provides its minimum realization), a cost that is based on a plainly admissible (and thus, without loss of generality, sub-optimal) input $\vect{u}_i \not= \overline{\vect{u}}_i^{\star}$ will result in a configuration where: $J_i \Big(\vect{e}_i(t), \vect{u}_i\big(\cdot;\ \vect{e}_i(t)\big)\Big)$ $\geq J_i \Big(\vect{e}_i(t), \overline{\vect{u}}_i^{\star}\big(\cdot;\ \vect{e}_i(t)\big)\Big)$.

Let us now begin our investigation on the sign of the difference between the cost
that results from the application of the feasible input $\vect{u}_{1,i}$,
which we shall denote by $\overline{J}_i\big(\vect{e}_i(t_{k+1})\big)$,
and the optimal cost $J_i^{\star}\big(\vect{e}_i(t_k)\big)$, while recalling that:
$J_i \big(\vect{e}_i(t), \overline{\vect{u}}_i (\cdot)\big)$ $=$
$\int_{t}^{t + T_p} F_i \big(\overline{\vect{e}}_i(s), \overline{\vect{u}}_i (s)\big) ds$ $+$
$V_i \big(\overline{\vect{e}}_i (t + T_p)\big)$:
\begin{align}
\overline{J}_i\big(\vect{e}_i(t_{k+1})\big) - J_i^{\star}\big(\vect{e}_i(t_k)\big) =\
& V_i \big(\vect{e}_{1,i} (t_{k+1} + T_p)\big) + \int_{t_{k+1}}^{t_{k+1} + T_p} F_i \big(\vect{e}_{1,i}(s), \vect{u}_{1,i} (s)\big) ds \notag \\
-&V_i \big(\vect{e}_{0,i} (t_k + T_p)\big) - \int_{t_k}^{t_k + T_p} F_i \big(\vect{e}_{0,i}(s), \vect{u}_{0,i} (s)\big) ds.
\end{align}
Considering that $t_k < t_{k+1} < t_k + T_p < t_{k+1} + T_p$, we break down the
two integrals above in between these integrals:
\begin{align}
&\overline{J}_i\big(\vect{e}_i(t_{k+1})\big) - J_i^{\star}\big(\vect{e}_i(t_k)\big) = \notag \\
V_i \big(\vect{e}_{1,i} (t_{k+1} + T_p)\big)
&\hspace{0mm}+ \int_{t_{k+1}}^{t_k + T_p} F_i \big(\vect{e}_{1,i}(s), \vect{u}_{1,i} (s)\big) d s
+ \int_{t_k + T_p}^{t_{k+1} + T_p} F_i \big(\vect{e}_{1,i}(s), \vect{u}_{1,i} (s)\big) d s \notag \\
-V_i \big(\vect{e}_{0,i} (t_k + T_p)\big)
&\hspace{0mm}- \int_{t_k}^{t_{k+1}} F_i \big(\vect{e}_{0,i}(s), \vect{u}_{0,i} (s)\big) d s
- \int_{t_{k+1}}^{t_k + T_p} F_i \big(\vect{e}_{0,i}(s), \vect{u}_{0,i} (s)\big) d s.
\label{eq:convergence_4_integrals_2}
\end{align}
\noindent Let us first focus on the difference between the two intervals in \eqref{eq:convergence_4_integrals_2} over $[t_{k+1}, t_{k+1} + T_p]$:
\begin{align}
& \int_{t_{k+1}}^{t_k  + T_p} F_i  \big(\vect{e}_{1,i}(s), \vect{u}_{1,i} (s)\big) d s
- \int_{t_{k+1}}^{t_k + T_p} F_i \big(\vect{e}_{0,i}(s), \vect{u}_{0,i} (s)\big) d s \notag \\
& \leq \bigg| \int_{t_k+h}^{t_k + T_p} F_i \big(\vect{e}_{1,i}(s), \vect{u}_{1,i} (s)\big) d s
- \int_{t_k+h}^{t_k + T_p} F_i \big(\vect{e}_{0,i}(s), \vect{u}_{0,i} (s)\big) d s \bigg| \notag \\
& \leq L_{F_i}\int_{h}^{T_p} \bigg\| \overline{\vect{e}}_i\big(t_k + s;\ \overline{\vect{u}}_i^{\star}(\cdot), \vect{e}_i(t_k + h) \big)
-  \overline{\vect{e}}_i\big(t_k + s;\ \overline{\vect{u}}_i^{\star}(\cdot), \vect{e}_i(t_k)\big) \bigg\| d s.
\label{eq:integrals_over_same_u_LV}
\end{align}

Consulting with Remark \ref{remark:predicted_actual_equations_with_disturbance} for the two different initial conditions we get: $\overline{\vect{e}}_i\big(t_k + s;\ \overline{\vect{u}}_i^{\star}(\cdot), \vect{e}_i(t_k +h)\big) = \vect{e}_i(t_k +h)$ $+ \int_{t_k +h}^{t_k + s} g_i\big(\overline{\vect{e}}_i(\tau;\ \vect{e}_i(t_k + h)), \overline{\vect{u}}_i^{\star}(\tau)\big) d\tau$, and $\overline{\vect{e}}_i\big(t_k + s;\ \overline{\vect{u}}_i^{\star}(\cdot), \vect{e}_i(t_k)\big) = \vect{e}_i(t_k)$ $+ \int_{t_k}^{t_k + h} g_i\big(\overline{\vect{e}}_i(\tau;\ \vect{e}_i(t_k)), \overline{\vect{u}}_i^{\star}(\tau)\big) d\tau$ $+ \int_{t_k + h}^{t_k + s} g_i\big(\overline{\vect{e}}_i(\tau;\ \vect{e}_i(t_k)), \overline{\vect{u}}_i^{\star}(\tau)\big) d\tau$.
Subtracting the latter from the former and taking norms on either side yields:
\begin{align}
&\bigg \| \overline{\vect{e}}_i\big(t_k + s;\ \overline{\vect{u}}_i^{\star}(\cdot), \vect{e}_i(t_k +h)\big)
-\overline{\vect{e}}_i\big(t_k + s;\ \overline{\vect{u}}_i^{\star}(\cdot), \vect{e}_i(t_k)\big) \bigg \| \leq \bigg\| \vect{e}_i(t_k +h) - \overline{\vect{e}}_i(t_k + h) \bigg \| \notag \\
&\hspace{17mm}+ L_{g_i} \int_{h}^{s} \bigg\| \overline{\vect{e}}_i\big(t_k + \tau;\ \overline{\vect{u}}_i^{\star}(\cdot), \vect{e}_i(t_k + h) \big)
-  \overline{\vect{e}}_i\big(t_k + \tau;\ \overline{\vect{u}}_i^{\star}(\cdot), \vect{e}_i(t_k)\big) \bigg\| d\tau.
\label{eq:df_interim_es}
\end{align}

\noindent By using Lemma \ref{lemma:diff_state_from_same_conditions} and applying the the Gr\"{o}nwall-Bellman inequality, \eqref{eq:df_interim_es} becomes: $\bigg \| \overline{\vect{e}}_i\big(t_k + s;\ \overline{\vect{u}}_i^{\star}(\cdot)$, $\vect{e}_i(t_k +h)\big)$ $-\overline{\vect{e}}_i\big(t_k + s;\ \overline{\vect{u}}_i^{\star}(\cdot)$, $\vect{e}_i(t_k)\big) \bigg \|$ $\leq \dfrac{\overline{w}_i}{L_{g_i}} (e^{L_{g_i}h} - 1) e^{L_{g_i}(s-h)}$.

\noindent Given the above result, \eqref{eq:integrals_over_same_u_LV} becomes:
\begin{align}
\int_{t_{k+1}}^{t_k + T_p} F_i \big(\vect{e}_{1,i}(s), \vect{u}_{1,i} (s)\big) d s
&- \int_{t_{k+1}}^{t_k + T_p} F_i \big(\vect{e}_{0,i}(s), \vect{u}_{0,i} (s)\big) d s \notag \\
& \leq L_{F_i} \dfrac{\overline{w}_i}{L_{g_i}^2} (e^{L_{g_i} h} - 1) (e^{L_{g_i}(T_p-h)} - 1).
\label{eq:end_result_two_integrals}
\end{align}
With this result established, we turn back to the remaining terms found in \eqref{eq:convergence_4_integrals_2} and, in particular, we focus on
the integral: $\int_{t_k + T_p}^{t_{k+1} + T_p} F_i \big(\vect{e}_{1,i}(s), \vect{u}_{1,i} (s)\big) d s$. We discern that the range of this integral has a length equal to the length of the interval where \eqref{eq:phi_psi} of Assumption \ref{ass:psi_psi} holds. Integrating \eqref{eq:phi_psi} over the interval $[t_k + T_p, t_{k+1} + T_p]$, for the controls and states applicable in it we get: $\int_{t_k + T_p}^{t_{k+1} + T_p} \Bigg(\dfrac{\partial V_i}{\partial \vect{e}_{1,i}} g_i\big(\vect{e}_{1,i}(s), \vect{u}_{1,i}(s)\big)$ $+ F_i\big(\vect{e}_{1,i}(s), \vect{u}_{1,i}(s)\big)\Bigg) ds \leq 0$ $\Leftrightarrow V_i\big(\vect{e}_{1,i}(t_{k+1} + T_p)\big)$
$+ \int_{t_k + T_p}^{t_{k+1} + T_p} F_i\big(\vect{e}_{1,i}(s), \vect{u}_{1,i}(s)\big) ds$ $\leq V_i\big(\vect{e}_{1,i}(t_k + T_p)\big)$. The left-hand side expression is the same as the first two terms in the right-hand side of equality \eqref{eq:convergence_4_integrals_2}. We can introduce the third one by subtracting it from both sides: $V_i\big(\vect{e}_{1,i}(t_{k+1} + T_p)\big)$ $+ \int_{t_k + T_p}^{t_{k+1} + T_p} F_i\big(\vect{e}_{1,i}(s), \vect{u}_{1,i}(s)\big) ds$
$- V_i\big(\vect{e}_{0,i}(t_k + T_p)\big)$ $\leq L_{V_i} \dfrac{\overline{w}_i}{L_{g_i}} (e^{L_{g_i}h} - 1) e^{L_{g_i} (T_p - h)}$.
Hence, we obtain:
\begin{align}
V_i\big(\vect{e}_{1,i}(t_{k+1} + T_p)\big)
& + \int_{t_k + T_p}^{t_{k+1} + T_p} F_i\big(\vect{e}_{1,i}(s), \vect{u}_{1,i}(s)\big) ds
- V_i\big(\vect{e}_{0,i}(t_k + T_p)\big) \notag \\
&\leq L_{V_i}\dfrac{\overline{w}_i}{L_{g_i}} (e^{L_{g_i}h} - 1) e^{L_{g_i} (T_p - h)}.
\label{eq:end_result_diff_V_plus_int}
\end{align}
Adding the inequalities \eqref{eq:end_result_two_integrals} and \eqref{eq:end_result_diff_V_plus_int} it is derived that: $\int_{t_{k+1}}^{t_k + T_p} F_i \big(\vect{e}_{1,i}(s), \vect{u}_{1,i} (s)\big) d s$ $- \int_{t_{k+1}}^{t_k + T_p} F_i \big(\vect{e}_{0,i}(s), \vect{u}_{0,i} (s)\big) d s$ $+ V_i\big(\vect{e}_{1,i}(t_{k+1} + T_p)\big)$ $+ \int_{t_k + T_p}^{t_{k+1} + T_p} F_i\big(\vect{e}_{1,i}(s), \vect{u}_{1,i}(s)\big) ds$ $- V_i\big(\vect{e}_{0,i}(t_k + T_p)\big)$ $\leq L_{F_i} \dfrac{\overline{w}_i}{L_{g_i}^2} (e^{L_{g_i} h} - 1) (e^{L_{g_i}(T_p-h)} - 1)$ $+ L_{V_i}\dfrac{\overline{w}_i}{L_{g_i}} (e^{L_{g_i}h} - 1) e^{L_{g_i} (T_p - h)}$, and therefore \eqref{eq:convergence_4_integrals_2}, by bringing the integral ranging from $t_k$ to $t_{k+1}$ to the left-hand side, becomes: $\overline{J}_i\big(\vect{e}_i(t_{k+1})\big)$ $- J_i^{\star}\big(\vect{e}_i(t_k)\big)$ $+ \int_{t_k}^{t_{k+1}} F_i \big(\vect{e}_{0,i}(s), \vect{u}_{0,i} (s)\big) d s$ $\leq L_{F_i} \dfrac{\overline{w}_i}{L_{g_i}^2} (e^{L_{g_i} h} - 1) (e^{L_{g_i}(T_p-h)} - 1)$ $+ L_{V_i}\dfrac{\overline{w}_i}{L_{g_i}} (e^{L_{g_i}h} - 1) e^{L_{g_i} (T_p - h)}$. By rearranging terms, the cost difference becomes bounded by: $\overline{J}_i\big(\vect{e}_i(t_{k+1})\big)$ $- J_i^{\star}\big(\vect{e}_i(t_k)\big)$ $\le \ \xi_i \overline{w}_i$ $-\int_{t_k}^{t_{k+1}} F_i \big(\vect{e}_{0,i}(s), \vect{u}_{0,i} (s)\big) d s$, where: $\xi_i \triangleq \dfrac{1}{L_{g_i}} \bigg(e^{L_{g_i}h} - 1\bigg)
\bigg[\big(L_{V_i} + \dfrac{L_{F_i}}{L_{g_i}}\big) \big(e^{L_{g_i}(T_p-h)}-1\big) + L_{V_i} \bigg] > 0$\text{, and} $\xi_i \overline{w}_i$ is the contribution of the bounded additive
disturbance $\vect{w}_i(t)$ to the nominal cost difference; $F_i$ is a positive-definite function as a sum of a positive-definite
$\vect{u}_i^\top \mat{R}_i \vect{u}_i$ and a positive semi-definite function
$\vect{e}_i^\top \mat{Q}_i \vect{e}_i$. If we denote by
$m_i \triangleq \lambda_{\min}(\mat{Q}_i, \mat{R}_i) \geq 0$ the minimum eigenvalue
between those of matrices $\mat{R}_i, \mat{Q}_i$, this means that: $F_i \big(\vect{e}_{0,i}(s), \vect{u}_{0,i} (s)\big) \geq m_i \|\vect{e}_{0,i}(s)\|^2$. By integrating the above between the interval of interest $[t_k, t_{k+1}]$ we get: $-\int_{t_k}^{t_{k+1}} F_i \big(\vect{e}_{0,i}(s), \vect{u}_{0,i} (s)\big)$ $\leq -m_i \int_{t_k}^{t_{k+1}} \| \overline{\vect{e}}_i(s;\ \overline{\vect{u}}_i^{\star}, \vect{e}_i(t_k)) \|^2 ds$. This means that the cost difference is upper-bounded by: $\overline{J}_i\big(\vect{e}_i(t_{k+1})\big)$ $- J_i^{\star}\big(\vect{e}_i(t_k)\big)$ $\leq \xi_i \overline{w}_i$ $-m_i \int_{t_k}^{t_{k+1}} \| \overline{\vect{e}}_i(s;\ \overline{\vect{u}}_i^{\star}(\cdot), \vect{e}_i(t_k)) \|^2 ds$,
and since the cost $\overline{J}_i\big(\vect{e}_i(t_{k+1})\big)$ is, in general, sub-optimal: $J_i^{\star}\big(\vect{e}_i(t_{k+1})\big) - \overline{J}_i\big(\vect{e}_i(t_{k+1})\big) \leq 0$: $J_i^{\star}\big(\vect{e}_i(t_{k+1})\big)$ $- J_i^{\star}\big(\vect{e}_i(t_k)\big)$ $\leq \xi_i \overline{w}_i$ $- m_i \int_{t_k}^{t_{k+1}} \| \overline{\vect{e}}_i(s;\ \overline{\vect{u}}_i^{\star}(\cdot), \vect{e}_i(t_k)) \|^2 ds$. Let $\Xi_i(\vect{e}_i) \triangleq J_i^{\star}(\vect{e}_i)$. Then, between consecutive times $t_k$ and $t_{k+1}$ when the FHOCP is solved, the last inequality reforms into:
\begin{align}
\Xi_i\big(\vect{e}_i(t_{k+1})\big) - \Xi_i\big(\vect{e}_i(t_k)\big)
&\leq \int_{t_k}^{t_{k+1}} \bigg( \dfrac{\xi_i}{h} \|\vect{w}_i(s)\|
-  m_i \| \overline{\vect{e}}_i(s;\ \overline{\vect{u}}_i^{\star}(\cdot), \vect{e}_i(t_k)) \|^2 \bigg) ds.
\label{eq:check_for_ISS_here_main_branch}
\end{align}
The functions $\sigma\big(\|\vect{w}_i\|\big) \triangleq \dfrac{\xi_i}{h} \|\vect{w}_i\|$ and $\alpha_3\big(\|\vect{e}_i\|\big) \triangleq m_i \|\vect{e}_i\|^2$ are class
$\mathcal{K}$ functions, and therefore, according to Lemma \ref{lemma:V_i_lower_upper_bounded}, $\Xi_i\big(\vect{e}_i\big)$ is an ISS Lyapunov function in $\mathcal{E}_i$. Given this fact, the closed-loop system is input-to-state stable in $\mathcal{E}_i$. Inevitably then, given Assumptions \ref{ass:psi_psi} and \ref{ass:psi_omega}, and condition $(3)$ of Theorem \ref{theorem:with_disturbances}, the closed-loop
trajectories for the error state of agent $i \in \mathcal{V}$ reach
the terminal set $\Omega_i$ for all
$\vect{w}_i(t)$ with $\|\vect{w}_i(t)\| \leq \overline{w}_i$, at some point $t = t^{\star} \geq 0$. Once inside
$\Omega_i$, the trajectory is trapped there because of the
implications\footnote{For more details, refer to the discussion after the
	declaration of Theorem $7.6$ in \cite{marquez2003nonlinear}.} of
\eqref{eq:check_for_ISS_here_main_branch} and Assumption \ref{ass:psi_omega}.

In turn, this means that the system \eqref{eq:error_system_perturbed} converges
to $\vect{z}_{i, \text{des}}$ and is trapped in a vicinity of it $-$ smaller than that
in which it would have been trapped (if actively trapped at all)
in the case of unattenuated disturbances $-$, while simultaneously
conforming to all constraints $\mathcal{Z}_{i}$. This conclusion holds
for all $i \in \mathcal{V}$, and hence, the overall system of agents
$\mathcal{V}$ is stable. \qed
\end{document}